\documentclass{article}
\usepackage{arxiv}

\pdfoutput=1

\usepackage{authblk}
\usepackage{hyperref}
\usepackage[utf8]{inputenc}
\usepackage[T1]{fontenc}
\usepackage{graphicx}

\usepackage{subfigure}

\usepackage[linesnumbered, vlined, ruled]{algorithm2e}
\usepackage{xcolor}
\usepackage{mathtools}
\usepackage{amsfonts}
\usepackage{amsmath}

\newtheorem{problem}{Problem}
\newtheorem{mydefinition}{Definition}
\newtheorem{mylemma}{Lemma}

\usepackage{soul,xcolor}
\setstcolor{red}
\date{}

\begin{document}

\title{Local Pair and Bundle Discovery over Co-Evolving Time Series}

\author[1]{Georgios Chatzigeorgakidis}
\author[2]{Dimitrios Skoutas}
\author[3]{Kostas Patroumpas}
\author[4]{Themis Palpanas}
\author[5]{Spiros Athanasiou}
\author[6]{Spiros Skiadopoulos}

\affil[1]{IMSI, Athena R.C., Greece, \textit{email: gchatzi@athenarc.gr}}
\affil[2]{IMSI, Athena R.C., Greece, \textit{email: dskoutas@athenarc.gr}}
\affil[3]{IMSI, Athena R.C., Greece, \textit{email: kpatro@athenarc.gr}}
\affil[4]{LIPADE, Paris Descartes University, France, French University Institute (IUF) \textit{email: themis@mi.parisdescartes.fr}}
\affil[5]{IMSI, Athena R.C., Greece, \textit{email: spathan@athenarc.gr}}
\affil[6]{Dept. of Inf. \& Telecommunications University of Peloponnese, Greece \textit{email: spiros@uop.gr}}

\maketitle

\begin{abstract}
    Time series exploration and mining has many applications across several industrial and scientific domains. In this paper, we consider the problem of detecting locally similar pairs and groups, called bundles, over co-evolving time series. These are pairs or groups of subsequences whose values do not differ by more than $\epsilon$ for at least $\delta$ consecutive timestamps, thus indicating common local patterns and trends. We first present a baseline algorithm that performs a sweep line scan across all timestamps to identify matches. Then, we propose a filter-verification technique that only examines candidate matches at judiciously chosen checkpoints across time. Specifically, we introduce two block scanning algorithms for discovering local pairs and bundles respectively, which leverage the potential of checkpoints to aggressively prune the search space. We experimentally evaluate our methods against real-world and synthetic datasets, demonstrating a speed-up in execution time by an order of magnitude over the baseline. This paper has been published in the 16\textsuperscript{th} International Symposium on Spatial and Temporal Databases (SSTD19).
\end{abstract}

\section{Introduction}
\label{sec:intro}
Time series are important in many applications both in industry, e.g. in energy and finance, as well as in science, e.g. in astronomy and biology~\cite{DBLP:journals/sigmod/Palpanas15}. Therefore, efficient management and mining of time series is a task of critical importance, but also highly challenging due to the large volume and complex nature of this data. 

Co-evolving time series are time series that are time-aligned, i.e. they contain observation values at the same timestamps all along their duration. In this work, we focus on discovering all \textit{pairs} or \textit{groups} (called \textit{bundles}) of \textit{locally} similar co-evolving time series and extracting the \textit{subsequences} where this local similarity occurs. We consider two time series as \textit{locally similar} if the pairwise distance of their values per timestamp is at most $\epsilon$ for a time interval that lasts at least $\delta$ \textit{consecutive} timestamps.

Several research efforts have focused on similarity search over time series to detect insightful patterns within a single or across a set of time series \cite{rakthanmanon2012searching, yeh2016matrix, linardi2018scalable, DBLP:conf/sofsem/Palpanas16}. However, to the best of our knowledge, the problem of discovering pairs or groups of similar \textit{time-aligned subsequences} within a set of \textit{co-evolving} time series has been overlooked.

Discovering such pairs and bundles is useful in various applications. For instance, public utility companies employ smart meters to collect time series measuring consumption per household (e.g., for water or electricity). Identifying such bundles of time series (i.e., a number of similar subsequences over certain time intervals) can reveal similar patterns of consumption among users, allowing for more personalized billing schemes. In finance, examining time series of stock prices can identify pairs or bundles of stocks trending similarly at competitive prices over some trading period (hours or days), hence offering precious insight for possible future investments.


\begin{figure}[b]
    \centering
    \includegraphics[width=0.75\columnwidth]{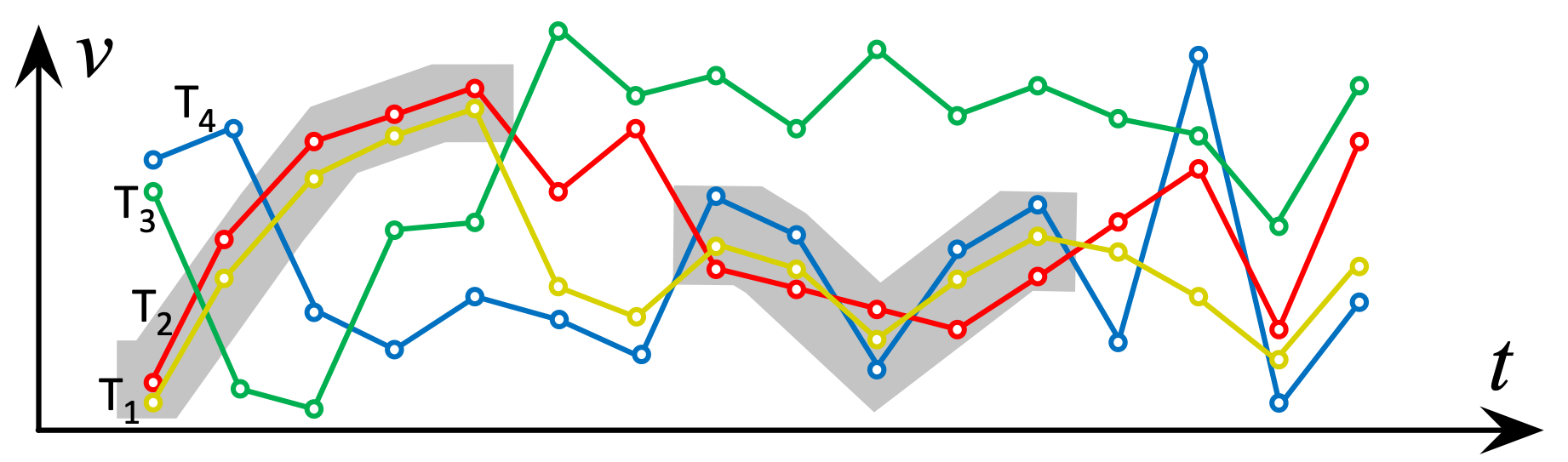}
    \caption{Example of a pair and a bundle of locally similar time series.}
    \label{fig:pair_flock_ex}
\end{figure}

Figure~\ref{fig:pair_flock_ex} illustrates an example comprising four time series depicted with different colors. We observe that from timestamp 1 to 5 the values of $T_1$ and $T_2$ are very close to each other, thus forming a locally similar pair. Similarly, from timestamp 8 to 12, the values of $T_1$, $T_2$ and $T_4$ are close to each other, forming a bundle with three members. Note that values in each qualifying subsequence may fluctuate along a bundle as long as they remain close to the respective values per timestamp of the other members in that bundle.

\begin{figure}[tb]
    \centering
    \includegraphics[width=0.75\columnwidth]{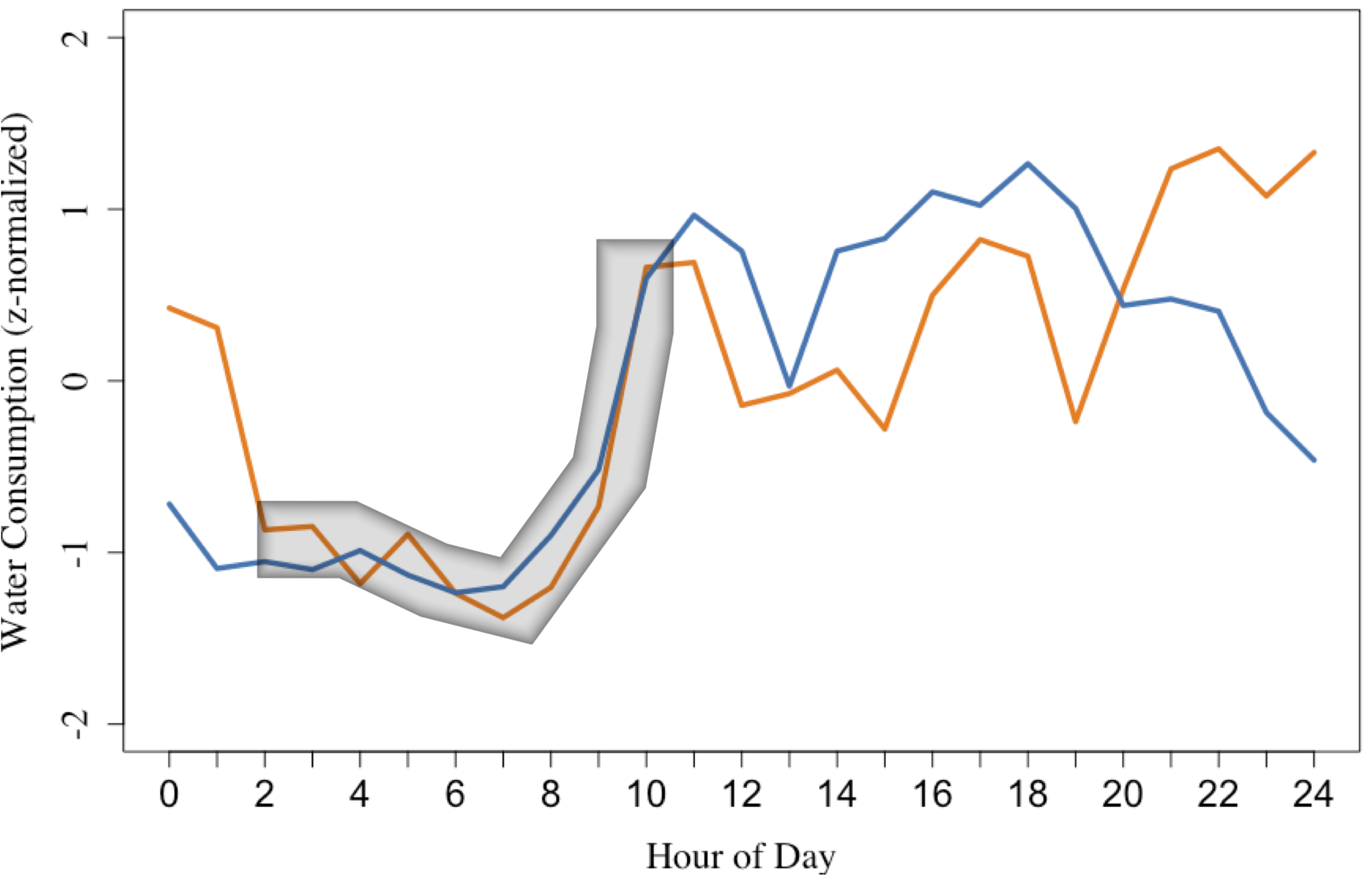}
    \caption{Example of a pair of locally (but not globally) similar time series.} 
    \label{fig:water_real_ex}
\end{figure}

A real-world example is depicted in Figure~\ref{fig:water_real_ex}. These two time series represent per-hour average water consumption during a day of the week for two different households. We can observe that their respective values per timestamp (at granularity of hours, in this example) are very close to each other during a certain time period (hours 2-11), but are farther apart in the rest. Hence, an algorithm that measures the global similarity between two time series might not consider this pair as similar; however, the subsequences inside the gray strip are clearly pairwise similar, and might indicate an interesting pattern. Identifying such local similarities within a sufficiently long time interval is our focus in this paper.


\begin{figure}[tb]
    \centering
    \includegraphics[width=0.75\columnwidth]{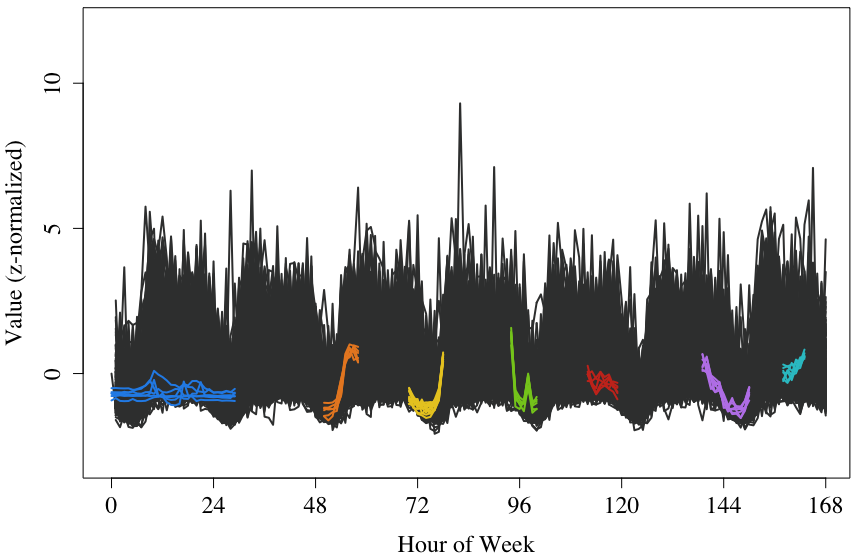}
    \caption{Discovered bundles of locally similar time series in a water consumption dataset.}
    \label{fig:bundles_water}
\end{figure}

Furthermore, Figure \ref{fig:bundles_water} depicts several bundles of locally similar time series detected by our algorithms in a real-world dataset containing smart water meter measurements. The detected bundles represent different per-hour average water consumption patterns during a week. There is a wider pattern detected among 6 households during the first 30 hours of the week indicating reduced consumption (probably no permanent residence). The orange and yellow patterns indicate different morning routines during the third and fourth day of the week. The green and purple patterns represent a reduction in consumption during the late hours of the fourth and sixth day, respectively, with some intermediate consumption taking place during the night. Finally, the shorter red and light blue bundles suggest different evening patterns for two other days (respectively, decreasing and increasing consumption).

Discovering all possible pairs and bundles of locally similar time series, along with the corresponding subsequences, within large sets is a computationally expensive process. To find matches, a filter-verification technique can be applied. At each timestamp, the {\em filtering} step can discover candidate pairs or groups having values close to each other; then, the {\em verification} step is invoked to determine whether each such candidate satisfies the required conditions, essentially whether this match occurs throughout a sufficiently large time interval. However, both the filtering and the verification steps are expensive. The computational cost becomes especially high for the case of bundle discovery, as it has to examine all possible subsets of locally similar time series that could form a bundle. Hence, such an exhaustive search is prohibitive when the number and/or the length of the time series is large.

In this paper, we employ a value discretization approach that divides the value axis in ranges equal to the value difference threshold $\epsilon$, in order to reduce the number of candidate pairs or bundles that need to be checked per timestamp. Leveraging this, we first propose two \textit{sweep line} scan algorithms, for pair and bundle discovery respectively, which operate according to the aforementioned filter-verification strategy. However, this process still incurs an excessive amount of comparisons, as it needs to scan all values at every timestamp. To overcome this, we introduce a more aggressive filtering  that only checks at selected \textit{checkpoints} across time, but ensuring that no false negatives ever occur. This approach incurs significant savings in computation cost, as we only need to examine candidate matches on those checkpoints only instead of all timestamps. To further reduce the number of examined candidates, we propose a strategy that judiciously places these checkpoints across the time axis in a more efficient manner. We then exploit these optimizations introducing two more efficient algorithms that significantly reduce the execution cost for both pair and bundle discovery.

The bundle discovery problem we address in this paper resembles the problem of \textit{flock discovery in moving objects}, where the goal is to identify sufficiently large groups of objects that move close to each other over a sufficiently long period of time \cite{gudmundsson2006computing,benkert2008reporting,vieira2009line,tanaka2015efficient}. In fact, the baseline algorithm we describe can be viewed as an adaptation of the algorithm presented  in \cite{vieira2009line}. However, to the best of our knowledge, ours is the first work to address the problems of locally similar pair and bundle discovery over co-evolving time series. Specifically, our main contributions can be summarized as follows:

\begin{itemize} 
 \item We introduce the problems of local pair and bundle discovery over co-evolving time series.
 \item We suggest an aggressive checkpoint-based pruning method that drastically reduces the candidate pairs and bundles that need to be verified, significantly improving performance.
 \item We conduct an extensive experimental evaluation using both real-world and synthetic time series, showing that our algorithms outperform the respective sweep line baselines.
\end{itemize}

The remainder of this paper is organized as follows. Section \ref{sec:related} reviews related work. Section \ref{sec:problem} describes the problems. Sections \ref{sec:local_join} and \ref{sec:bundle_disc} introduce our algorithms for pair and bundle discovery, respectively. Section \ref{sec:eval} reports our experimental results, and finally Section \ref{sec:conclusions} concludes the paper.


\section{Related Work}
\label{sec:related}

\textbf{Time series similarity.} 
Similarity search over time series has attracted a lot of research interest~\cite{DBLP:journals/pvldb/EchihabiZPB18}. One well-studied family of approaches includes wavelet-based methods~\cite{chan1999icde}, which rely on \emph{Discrete Wavelet Transform}~\cite{graps1995cse} to reduce the dimensionality of time series and generate an index using the coefficients of the transformed sequences. The \emph{Symbolic Aggregate Approximation} (SAX) representation~\cite{jessica2007dmkd} has led to the design of a series of indices, including $i$SAX~\cite{shieh2008kdd}, $i$SAX 2.0~\cite{camerra2010icdm}, $i$SAX2+~\cite{camerra2014kais}, ADS+~\cite{zoumpatianos2014sigmod}, Coconut~\cite{DBLP:journals/pvldb/KondylakisDZP18}, DPiSAX~\cite{dpisaxjournal}, and ParIS~\cite{DBLP:conf/bigdataconf/PengFP18}. However, these indices support similarity search over complete time series, i.e. whole-matching. Recently, the \textit{ULISSE} index was proposed~\cite{linardi2018scalable}, which is the first index that can answer similarity search queries of variable length. 


Moreover, many approaches have been proposed for subsequence matching. In this problem, a query subsequence is provided and the goal is to identify matches of this subsequence across one or more time series, typically of large length. The \textit{UCR suite} \cite{rakthanmanon2012searching} offers a framework comprising four different optimizations regarding subsequence similarity search. In computing \textit{full-similarity-joins} over large collections of time series, i.e., to detect for each possible subsequence its \textit{nearest neighbor}, the \textit{matrix profile}~\cite{yeh2016matrix} keeps track of Euclidean distances among each pair within a \textit{similarity join set} (i.e., a set containing pairs of each subsequence with its nearest neighbor).

The problem we address in this paper differs from the above settings. Instead of identifying matches of a query subsequence against one, or more time series, we are interested in discovering locally similar pairs and bundles of time-aligned subsequences within a given collection of time series.



\textbf{Time series clustering.}
Our work also relates to {\em clustering of time series}, where methods perform either partitioning or density-based clustering. In the former class, algorithms typically partition the time series into $k$ clusters. Similarly to iterative refinement employed in $k$-means, the $k$-Shape partitioning algorithm~\cite{Paparrizos:2015:KEA:2723372.2737793,Paparrizos:2017:FAT:3086510.3044711} aims to preserve the shapes of time series assigned to each cluster by considering the shape-based distance, a normalized version of the cross-correlation measure between time series. In contrast, density-based clustering methods are able to identify clusters of time series with arbitrary shapes. YADING~\cite{Ding:2015:YFC:2735479.2735481} is a highly efficient and accurate such algorithm, which consists of three steps: it first samples the input time series also employing PAA (Piecewise Aggregate Approximation) to reduce the dimensionality, then applies multi-density clustering over the samples, and finally assigns the rest of the input to the identified clusters. However, clustering methods consider time series in their entirety and not matching subsequences as we consider in this work.

\textbf{Discovery of movement patterns in trajectories.}
Our work also relates to approaches for discovering clusters of moving objects, in particular a type of movement patterns that is referred to as {\em flocks}~\cite{gudmundsson2006computing}. A flock is a group of at least $m$ objects moving together within a circular disk of diameter $\epsilon$ for at least $\delta$ consecutive timestamps. Finding an exact flock is NP-hard, hence this work suggests an \textit{approximate} solution to find the \textit{maximal} flock from a set of trajectories using computational geometry concepts. In \cite{benkert2008reporting}, another \textit{approximate} solution for detecting all flocks is based on a skip-quadtree that indexes sub-trajectories. Flock discovery over {\em streaming} positions from moving objects was addressed in \cite{vieira2009line}. This \textit{exact} solution discovers flock disks that cover a set of points at each timestamp. Their flock discovery algorithm finds candidate flocks per timestamp and joins them with the candidate ones from the previous timestamps, reporting a flock as a result when it exceeds the time constraint $\delta$. An improvement over this technique was presented in \cite{tanaka2015efficient}, using a \textit{plane sweeping} technique to accelerate detection of object candidates per flock at each timestamp, while an inverted index speeds up comparisons between candidate disks across time. In our setting, detection of bundles is similar to flocks, thus for our baseline method we adapt the algorithm from \cite{vieira2009line}.
\section{Problem Definition}
\label{sec:problem}


A {\em time series} is a time-ordered sequence $X = \{X^1, X^2 \ldots, X^{k}\}$, where $X^{i}$ is the value at the $i$-th timestamp and $k$ is the length of the series (i.e., the number of timestamps). We consider a set of {\em co-evolving} time series, so all time series are {\em time-aligned} and each series has a value at each of the $k$ timestamps. Given a set of such co-evolving time series, our goal is to find {\em pairs} of time series that have similar values locally over some time intervals of significant duration. More specifically:

\begin{mydefinition}[Locally Similar Time Series]
Two co-evolving time series $X_i$ and $X_j$ are locally similar if there exists a time interval $I$ spanning at least $\delta$ consecutive timestamps such that at every timestamp in $I$ their corresponding values do not differ by more than a given threshold $\epsilon$, i.e., $\forall t \in I, |X^{t}_i - X^{t}_j| \leq \epsilon$.
\end{mydefinition}



Note that threshold $\epsilon$ expresses the maximum tolerable deviation per timestamp between two time series, so it actually concerns the absolute difference of their corresponding values. We wish to find all such pairs of time series, so the problem is actually a {\em self-join} over the dataset, specifying as join criteria the distance threshold $\epsilon$ and the minimum time duration $\delta$ of qualifying pairs. More formally:

\begin{problem}[Pair Discovery over Time Series]
Given a set of $n$ co-evolving time series $\mathcal{X}=\{X_1,X_2,...,X_n\}$ of equal duration $k$, a distance threshold $\epsilon>0$, and a time duration threshold $\delta>1$ timestamps, $(\delta \in N)$, retrieve all pairs $\{X_i, X_j\}, 1 \leq i < j \leq n$ of locally similar time series along with the corresponding time intervals.
\end{problem}

\begin{figure}[tb]
    \centering
    \includegraphics[width=0.75\columnwidth]{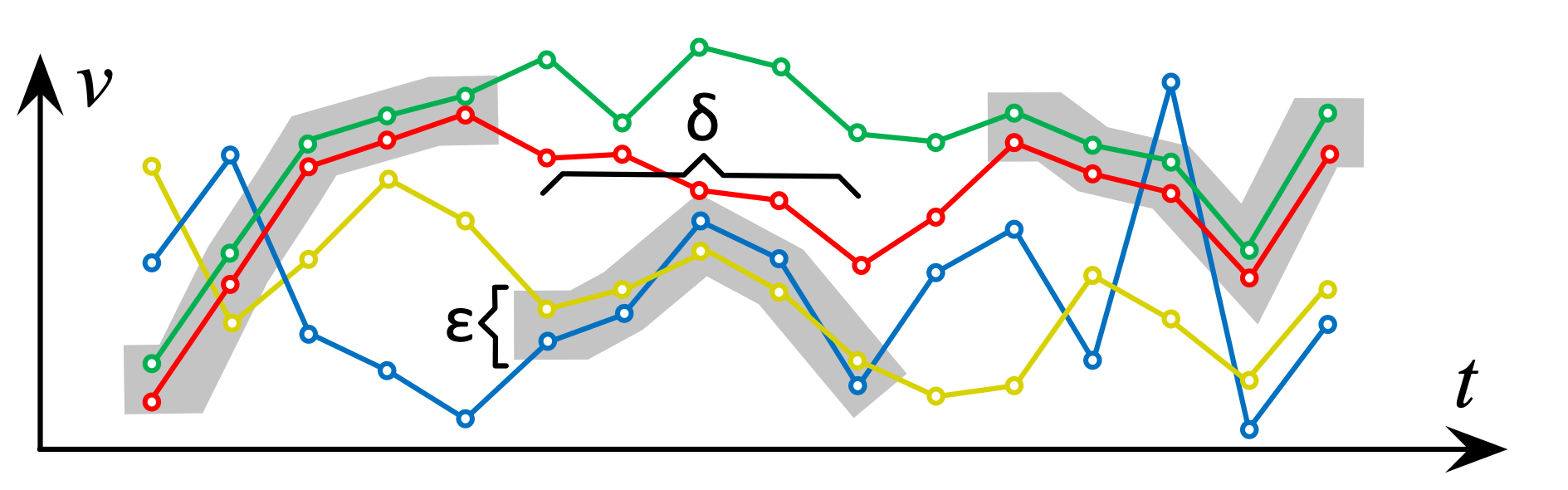}
    \caption{Pair discovery over a set of time series.}
    \label{fig:sim_join}
\end{figure}

For example, in Figure~\ref{fig:sim_join}, the detected pairs for a specified $\epsilon$ and duration $\delta$ would be the locally similar time series within the grey ribbons. Since two time series might be locally similar in more than one intervals, their matching subsequences are considered as two different pairs, one for each interval. For instance, in Figure~\ref{fig:sim_join} the green and red time series yield two matching pairs in different time intervals.

The above problem can be extended to the detection of groups, also called \textit{bundles}, of co-evolving time series. Instead of pairs, each such bundle of time series contains at least a pre-defined number $\mu > 2$ of members, which are pairwise locally similar to each other over a time interval of sufficient duration. This problem can be formulated as follows:

\begin{problem}[Local Bundle Discovery over Time Series]
Given a set of co-evolving time series $\mathcal{X}=\{X_1,X_2,...,X_n\}$ of equal length $k$, a minimum bundle size $\mu > 2, (\mu \in \mathbb{N}$), a maximum value difference $\epsilon>0$, and a minimum time duration $\delta>1$ timestamps, $(\delta \in \mathbb{N})$, retrieve all groups $\mathcal{G}$ of time series such that:

\begin{itemize}
    \item Each group $G \in \mathcal{G}$ contains at least $\mu$ time series.
    \item Within each group $G \in \mathcal{G}$, all pairs of time series are locally similar with respect to $\epsilon$ and $\delta$.
    \item Each group $G \in \mathcal{G}$ is \textit{maximal}, i.e., there is no other group $G' \supseteq G$ that also forms a bundle for the same time interval. 
\end{itemize}
\end{problem}

\begin{figure}[tb]
    \centering
    \includegraphics[width=0.75\columnwidth]{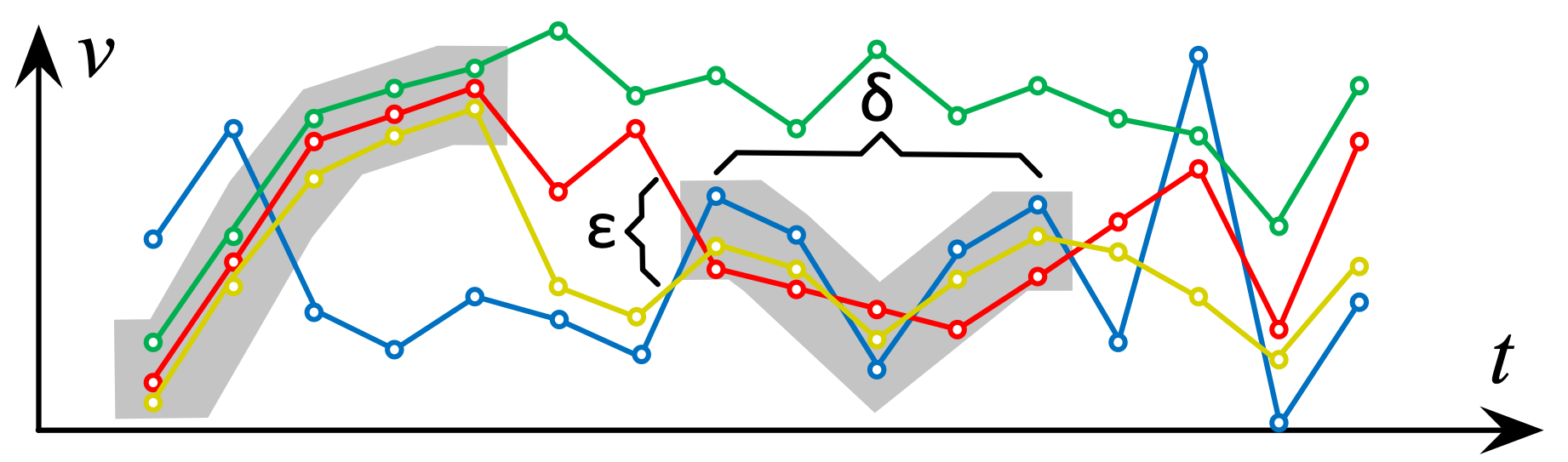}
    \caption{Bundle discovery over a set of time series.}
    \label{fig:bundle_disc}
\end{figure}

An illustration of the above problem is shown in Figure \ref{fig:bundle_disc}. Each grey band covers the subsequences of at least $\mu=3$ time series that constitute a bundle. These subsequences are pairwise locally similar for a specified distance $\epsilon$ and duration $\delta$.

\section{Pair Discovery}
\label{sec:local_join}
In this Section, we propose two solutions for the pair discovery problem. The first (Section~\ref{subsec:sweep_line_join}) is a baseline algorithm that uses a sweep line to scan the co-evolving time series throughout their duration, while validating and keeping all the pairs that satisfy the given constraints employing a value discretization scheme per timestamp (Section~\ref{subsec:indexing}). The second method (Section~\ref{subsec:checkpoint_join}) employs an optimization that reduces the number of pairs to consider by judiciously probing candidates at selected timestamp values (referred to as \textit{checkpoints}, Section~\ref{sec:checkpoint}). This significantly prunes the search space without missing any qualifying results.

\subsection{Value Discretization}
\label{subsec:indexing}
To reduce the candidate pairs that need be checked at each timestamp $t$, we discretize the values of all time series at $t$ in {\em bins}, i.e., several consecutive value ranges, each one of size $\epsilon$. Time series with values within the same bin at timestamp $t$ form candidate pairs, but we also need to check adjacent bins for additional candidate pairs whose values differ by at most $\epsilon$. Time series having values at non-adjacent bins are certainly farther than $\epsilon$ at that specific timestamp $t$, so we can avoid these checks. 

\begin{figure}[tb]
    \centering
    \includegraphics[width=0.25\columnwidth]{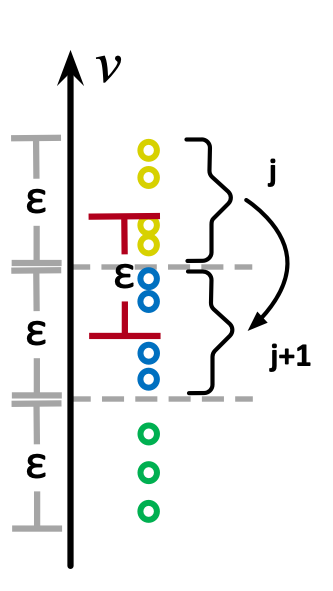}
    \caption{Discretization of time series values at timestamp $t$.}
    \label{fig:index}
\end{figure}

To detect all candidate pairs and avoid cross-checking in every two adjacent bins we consider a \textit{value range} of size $\epsilon$, whose upper endpoint coincides with each value under consideration at time $t$. Then, all values of time series contained within this range, form candidate pairs (see Figure~\ref{fig:index}). Obviously, values contained in the same bin $j$ can form candidate pairs. Then, we can cross-check each value in bin $j$ with values in bin $j+1$ for additional candidates with value difference at most $\epsilon$, as indicated with the red (right) range.

Thus, at each timestamp $t$, the process of finding all the pairs consists of the following two steps: (1) \textit{Filtering} - Search among time series values in adjacent bins to detect candidate pairs using the aforementioned search method. (2) \textit{Verification} - For each candidate pair of time series, check similarity of their respective values at successive timestamps as long as this pair still qualifies to the matching conditions (or the end of time series data is reached). This step actually resembles to a kind of ``horizontal expansion'' along the time axis in an attempt to eagerly verify and report pairs.

\subsection{Pair Discovery Using Sweep Line}
\label{subsec:sweep_line_join}
A baseline method for performing the pair discovery over a set of time series is to check all the candidate pairs formed at each timestamp, and verify whether the minimum duration constraint $\delta$ is satisfied. Algorithm \ref{alg:simple_scan_bundle} describes this procedure. Pair discovery (Line 5) considers a time duration $T$ (as a set of consecutive timestamps) to check for results. Initially $T = \{0, \dots, k\}$, where $k$ is the total duration of all time series data. For each timestamp $t$, we obtain only the subset of time series whose values are contained in two adjacent bins (Line 7). Based on these values at $t$, we obtain all candidate pairs with respect to the threshold $\epsilon$ (Line 9). For each such pair, if it is not already part of the resulting pairs at that specific timestamp $t$, we \textit{verify} it by first expanding it horizontally (Lines 10-12) and checking whether this pair meets the duration constraint $\delta$ (Line 13) along subsequent timestamps. If so, we add it to the reported results (Line 14).

Concerning the \textit{horizontal expansion}, we iterate over the subsequent timestamps (after the current one -- Line 17) and we stop when the threshold $\epsilon$ is violated (Lines 18--19). Afterwards, we mark the start of this pair with the current timestamp $t$, whereas its end is marked by the timestamp at which the threshold $\epsilon$ is crossed, and we return this pair (Lines 20-22).

\SetKwInOut{Parameter}{Parameters}

\setlength{\textfloatsep}{4pt}
\begin{algorithm}[tb]
	\DontPrintSemicolon
	\KwIn{Set $\mathcal{X}$ of co-evolving time series of length $k$}
	\Parameter{Threshold $\epsilon$, min duration $\delta$}
	\KwOut{List $P$ with all locally similar pairs of time series}
	\vspace{4pt}
	
	\begin{footnotesize}
	$B \leftarrow CreateBins(\mathcal{X})$ \\
	$P \leftarrow DiscoverPairs(\emptyset, B, \{0, ..., k\}, \epsilon, \delta)$ \\
	\KwRet $P$
	
	\vspace{4pt}
	\SetKwProg{procFilter}{Procedure}{}{}
	\procFilter{$DiscoverPairs(P, B, T, \epsilon, \delta)$}{
    	\ForEach{$t \in T$}{
            \ForEach{$z=0 \to B.size$}{
                $X \leftarrow B_z \cup B_{z+1}$ \\
                \ForEach{$X \in \mathcal{X}^t$}{
        	        $\mathcal{P}^t \leftarrow getAdjacentPairs(X^t, \epsilon)$ \\
        	        \ForEach{$p \in \mathcal{P}^t$} {
        	            \If{$p \notin P$}{
        	                $p \leftarrow VerifyPair(p, t, k, \epsilon)$ \\
        	                \If{$p.end-p.start \geq \delta$}{
        	                    $P \leftarrow P \cup p$ \\
        	                }
        	            }
        	        }
        	    }
    	    }
    	}
    	\KwRet $P$ \\
	}
	
	\vspace{4pt}
	\SetKwProg{procExpand}{Procedure}{}{}
	\procExpand{$VerifyPair(p, t, k, \epsilon)$}{
	    \ForEach{$t'=t+1 \to k$}{
	        \If{$|p.X_i^{t'}-p.X_j^{t'}| > \epsilon$}{
	            $break$ \\
	        }
	    }
	    $p.start \leftarrow t$ \\ 
	    $p.end \leftarrow t'$ \\ 
		\KwRet $p$ \\
	}
	\end{footnotesize}
	\caption{Sweep line scan pair discovery}
	\label{alg:simple_scan_sim}
\end{algorithm}

However, searching over all timestamps in such an exhaustive manner can be expensive, particularly for long time series. Next, we present an optimization that identifies candidate pairs at selected timestamps only, so that only those pairs require verification.

\subsection{Optimized Filtering at Checkpoints}
\label{sec:checkpoint}

\begin{figure}[tb]
    \centering
    \includegraphics[width=0.75\columnwidth]{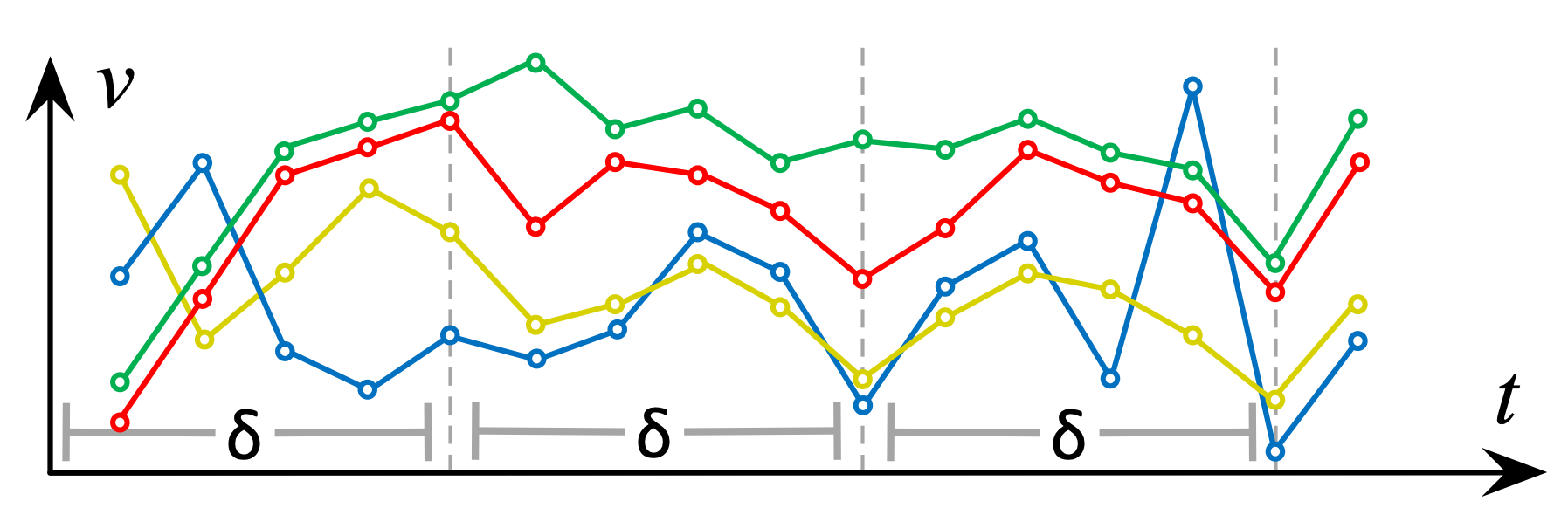}
    \caption{Checkpoints placed every $\delta$ timestamps.}
    \label{fig:checkpoints}
\end{figure}

To prune the search space, we consider \textit{checkpoints} along the time axis, so that searching for candidate pairs will be performed at these specific timestamps only. If the temporal span between two successive checkpoints does not exceed the minimal duration threshold $\delta$, we can ensure no false negatives, since any qualifying pair starting at an intermediate timestamp between two checkpoints will surely be detected at least on the second one. Figure \ref{fig:checkpoints} shows an example of a set of time series with checkpoints placed along the time axis every $\delta=5$ timestamps.

\begin{figure}[!t]
 \centering
 \subfigure[Pair with a starting point before $t'$.]{\includegraphics[width=0.45\textwidth]{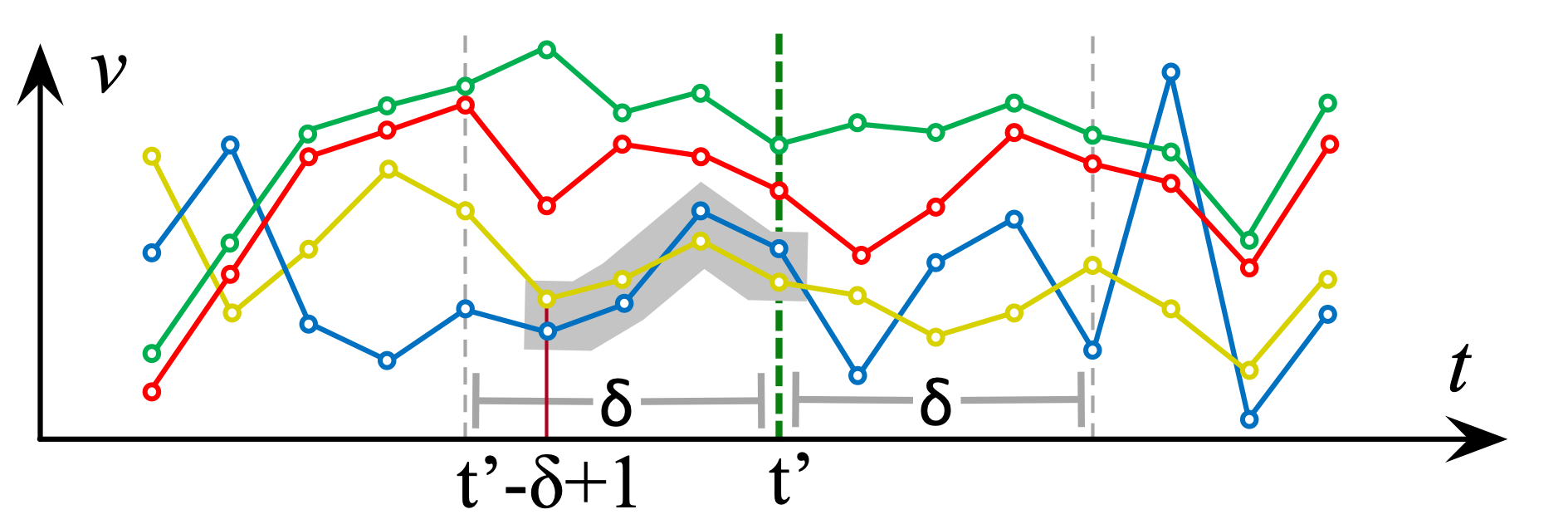}\label{fig:proof1}}
 \subfigure[Pair with an ending point after $t'$.]{\includegraphics[width=0.45\textwidth]{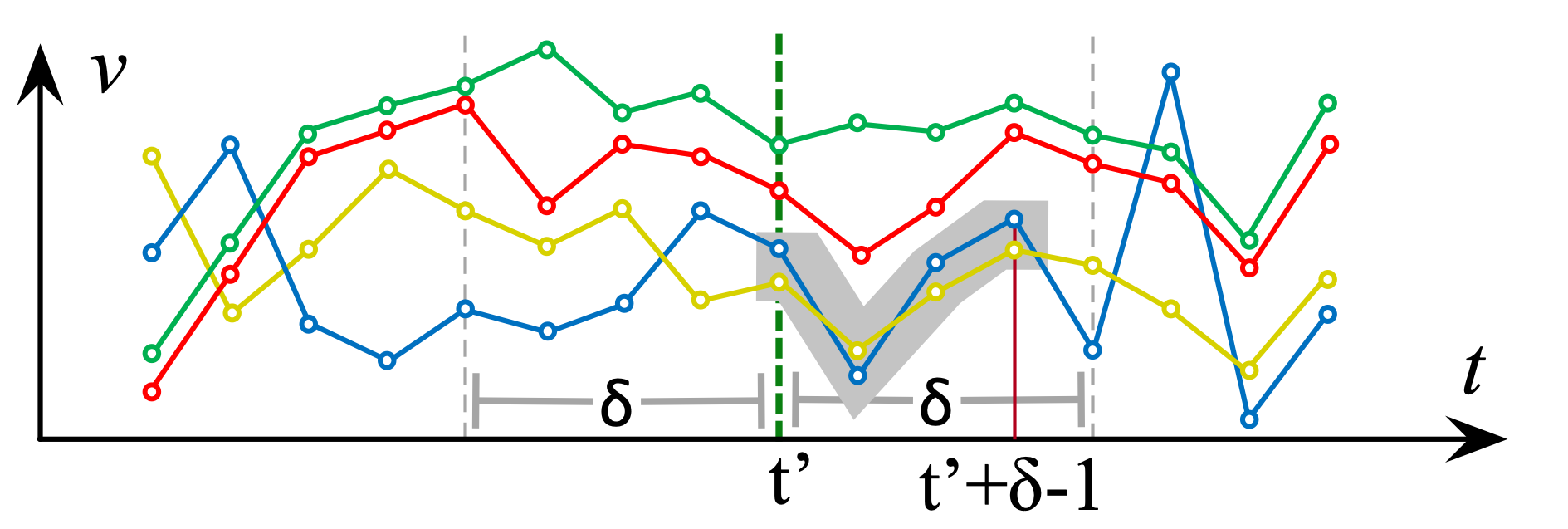}\label{fig:proof2}}
 \vspace{-10pt} 
 \caption{A qualifying pair will be detected on a checkpoint.}
 \label{fig:proof}
\end{figure}

Assume a set of checkpoints placed at time interval $\delta$ from each other, as depicted in Figure \ref{fig:proof1}. Let a checkpoint at timestamp $t'$ and a qualifying pair of duration $\delta$ starting at timestamp $t'-\delta+1$. This pair cannot have smaller duration, otherwise it would not meet constraint $\delta$. Consequently, the pair will be detectable on the checkpoint at $t'$, as shown in the figure. Similarly, if a qualifying pair ends at timestamp $t'+\delta-1$ (Figure \ref{fig:proof2}), it will be detected at the checkpoint at $t'$. Hence, all pairs around a checkpoint at $t'$ can be detected as candidates when we check their values at $t'$. Thus, we can easily conclude to the following observation.

\begin{mylemma}[Checkpoint Covering Interval]
Let the interval between successive checkpoints not exceed $\delta$. Considering a checkpoint placed at timestamp $t'$, all qualifying pairs starting at $s, t'-\delta+1 \leq s < t'$ and ending at $f, t' < f \leq t'+\delta-1$ will satisfy all matching constraints at timestamp $t'$.
\end{mylemma}

This lemma entails that it suffices to check for candidate pairs only at checkpoints, i.e., every $\delta$ timestamps. We denote the set of checkpoints as $C$.
Since we skip timestamps and in order to avoid false misses, we now have to verify pairs with a horizontal expansion (as in \ref{subsec:sweep_line_join}), but towards both directions, i.e., before and after a given checkpoint. Overall, at each checkpoint the optimized process performs: (1) \textit{Filtering} - Search among the values of time series in adjacent bins to detect candidate pairs. (2) \textit{Verification} - For each candidate pair, perform a \textit{two-way} horizontal expansion across the time axis.

\begin{figure}[tb]
    \centering
    \includegraphics[width=0.75\columnwidth]{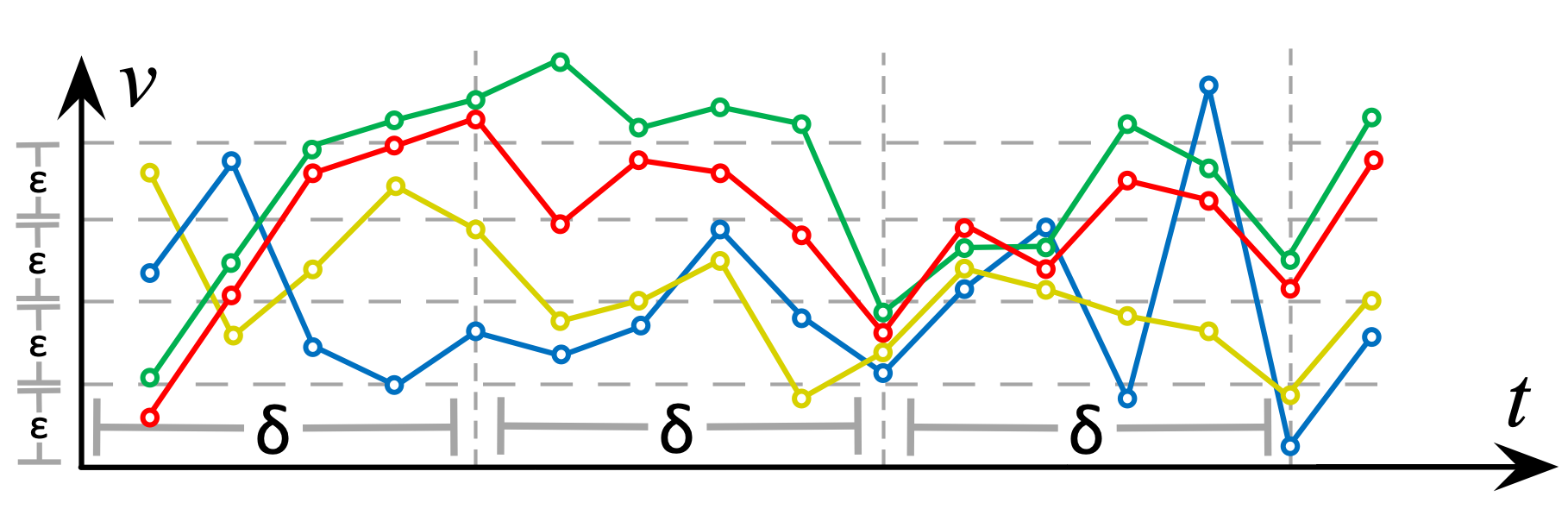}
    \caption{Sub-optimal checkpoint placement.}
    \label{fig:opt1}
\end{figure}
\begin{figure}[tb]
    \centering
    \includegraphics[width=0.75\columnwidth]{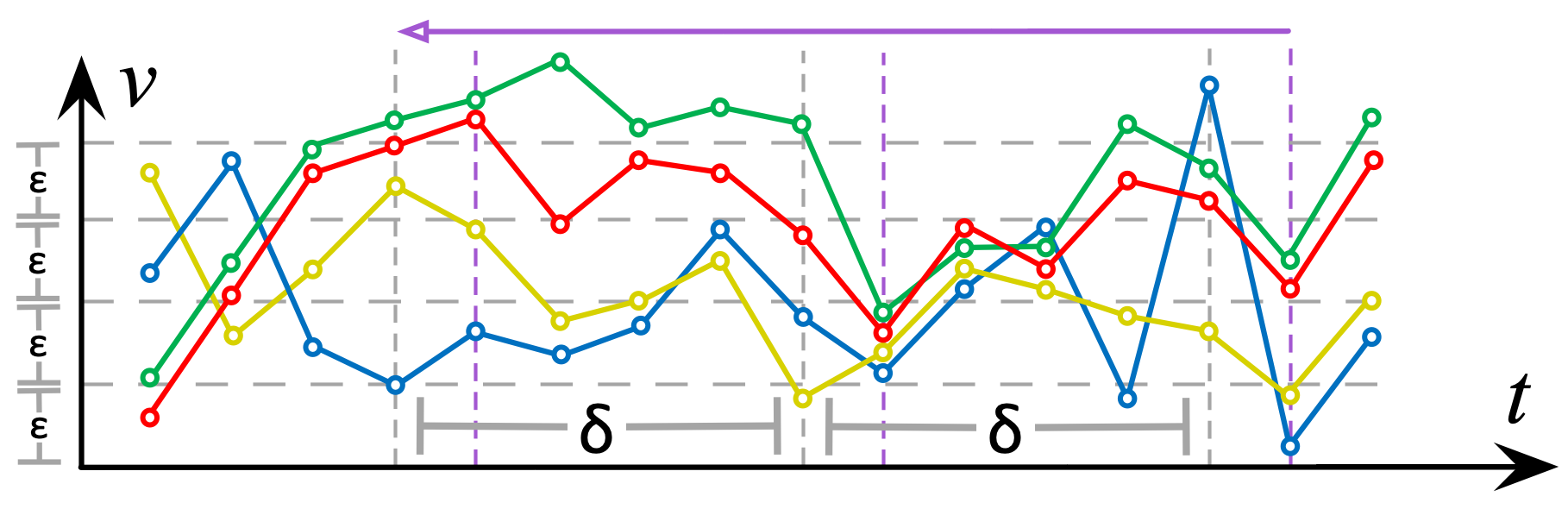}
    \caption{Improved checkpoint placement.}
    \label{fig:opt2}
\end{figure}

\paragraph{Improving placement of checkpoints} Depending on the dataset, the default checkpoint placement might yield an \textit{increased} number of candidate pairs that would incur too many (and perhaps unnecessary) verifications. Intuitively, if the time series values at a specific timestamp $t'$ are placed in a more \textit{``scattered''} manner over the bins, then less candidates would be generated. This is because the values of time series at $t'$ would differ from each other by more than $\epsilon$ and thus can be pruned as described in Section \ref{subsec:indexing}. Figure \ref{fig:opt1} depicts such a case of sub-optimal placement of checkpoints, where the second checkpoint is placed at a rather \textit{dense} area and as a result, six candidate pairs are considered. We can remedy this issue by shifting all checkpoints together either to the left or to the right, yet maintaining their temporal span every $\delta$. As shown in Figure \ref{fig:opt2}, all three checkpoints are collectively shifted to the left, avoiding the dense area and reducing the total number of candidate pairs. An extra checkpoint can be inserted before the first or after the last one, so as to guarantee that there is no interval longer than $\delta$ without checkpoints.


\begin{figure}[tb]
    \centering
    \includegraphics[width=0.75\columnwidth]{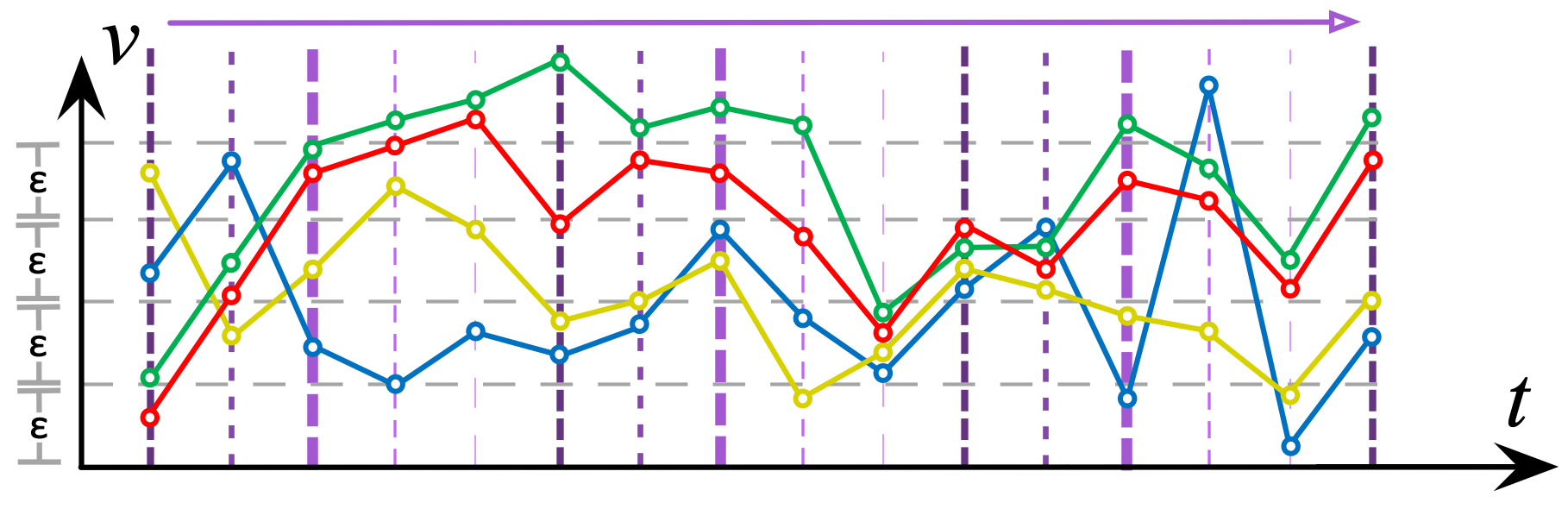}
    \caption{Best checkpoint placement (shown as thick vertical lines).}
    \label{fig:heuristic1}
\end{figure}

Clearly, the placement of the set $C$ of checkpoints influences the amount of candidate pairs. We wish to find the best such placement, which provides the least number of candidate pairs. The amount of candidates depends on the cardinality of the bins (i.e., the number of values in each one, as shown in Fig.~\ref{fig:index}) at any particular checkpoint $c \in C$. Given that $n$ is the total number of time series in the dataset (and hence the number of values at each checkpoint), we can identify the most populated bin at checkpoint $c$ by calculating the maximal density $\frac{\max\{B_{c}\}}{n}$, where $B_{c}$ represents the set of bin cardinalities at checkpoint $c$. Therefore, for a given configuration $C$ of checkpoints, we can estimate an overall cost $r$ by taking the sum of such maximal densities over all checkpoints, i.e.,  $r=\sum_{c \in C}{\frac{\max\{B_{c}\}}{n}}$. The less this total cost, the smaller the cardinality per bin at each checkpoint and, thus, the less the candidates that will be generated. Consequently, we seek to find the minimum $r$. To do so, we shift all checkpoints together to the right, one timestamp at a time, we estimate ratio $r$ again and repeat $\delta$ times. This procedure is illustrated in Figure \ref{fig:heuristic1}, where the checkpoints symbolized with similar lines belong to the same set as we move them to the right in order to identify the best placement (indicated with the thickest vertical dashed lines).

\setlength{\textfloatsep}{4pt}
\begin{algorithm}[tb]
	\DontPrintSemicolon
	\KwIn{Set $\mathcal{X}$ of co-evolving time series of length $k$}
	\Parameter{Threshold $\epsilon$, min duration $\delta$}
	\KwOut{A list $P$ containing all the locally similar time series}
	\vspace{4pt}
	
	\begin{footnotesize}
	$B \leftarrow CreateBins(\mathcal{X})$ \\
	$C \leftarrow GetCheckpoints(k, \delta)$ \\
	$P \leftarrow DiscoverPairs(\emptyset, B, C, \epsilon, \delta)$ \\
	\KwRet $P$ \\
	
	\vspace{4pt}
	\SetKwProg{procVerify}{Procedure}{}{}
	\procVerify{$VerifyPair2Way(p, t, k, \epsilon)$}{
	    \ForEach{$t'=t-1 \to 0$}{
	        \If{$|p.X_i^{t'}-p.X_j^{t'}| > \epsilon$}{
	            $break$ \\
	        }
	    }
	    $p.start \leftarrow t'$ \\
	    $t_c \leftarrow p.start + \delta - 1$ \\
	    \ForEach{$t'=t_c \to t$}{
	        \If{$|p.X_i^{t'}-p.X_j^{t'}| > \epsilon$}{
	            $p.f \leftarrow t$\\
	            \KwRet $p$ \\
	        }
	    }
	    \ForEach{$t'=t_c+1 \to k$}{
	        \If{$|p.X_i^{t'}-p.X_j^{t'}| > \epsilon$}{
	            $break$ \\
	        }
	    }
	    $p.end \leftarrow t'$ \\
		\KwRet $p$ \\
	}
	\end{footnotesize}
	\caption{Checkpoint scan pair discovery}
	\label{alg:checkpoint_scan_sim}
\end{algorithm}

\subsection{Pair Discovery Using Checkpoints}
\label{subsec:checkpoint_join}
After identifying the best checkpoint placement, we can discover pairs of locally similar time series by applying the exhaustive algorithm presented in Section~\ref{subsec:sweep_line_join}, but iterating over the defined checkpoints instead of all timestamps. To speed up the verification step, we introduce an optimization that reduces the number of checks. For each candidate pair $p$ that started at (a possibly previous) timestamp $p.start$, we first expand its verification to the left. In case the current duration of pair $p$ is still less than $\delta$, we jump at timestamp $p.start + \delta$ in order to eagerly prune candidates that will not last at least $\delta$. There, we check directly whether the values of these two time series qualify, and we continue this check backwards in time. If at an intermediate timestamp the $\epsilon$ constraint is not met, we can stop the verification process and discard the candidate pair.

The procedure for pair discovery is listed in Algorithm \ref{alg:checkpoint_scan_sim}. Initially, we calculate the best possible checkpoint set (Line 2). Then, we run the procedure described in Section \ref{subsec:sweep_line_join} but instead of probing over all timestamps we iterate over the resulting checkpoint set (Line 3).

Regarding verification, we first move towards the leftmost endpoint (Line 6) and detect the starting timestamp of the pair (Line 9). Then, we iterate from the timestamp $t_c = p.start + \delta$ towards the initial timestamp $t$ (Line 11). If the pair does not qualify at a timestamp during this interval (Line 12), we set $t$ as its final timestamp and return the pair (Line 14). Otherwise, we continue from timestamp $t_c$ and towards the rightmost timestamp until the pair ceases to qualify (Lines 15-19).

\section{Bundle Discovery}
\label{sec:bundle_disc}
We now consider the bundle discovery problem. We propose two algorithms: an exhaustive one using a sweep line (Section~\ref{subsec:sweepline_bundle}), and one using checkpoints (Section~\ref{sec:checkpoint_bundle}), following the same principles as in pair discovery. 


\subsection{Bundle Discovery Using Sweep Line}
\label{subsec:sweepline_bundle}
To exhaustively detect bundles of time series, we can follow a similar procedure employing a sweep line as in Section \ref{subsec:sweep_line_join}. However, this time we detect candidate bundles at each timestamp before verifying whether constraints concerning minimal duration $\delta$ and minimum membership $\mu$ are satisfied. Essentially, this can be thought of as an adaptation of the flock discovery approach in \cite{vieira2009line} to the 1-dimensional setting in the case of time series.


Algorithm \ref{alg:simple_scan_bundle} describes this exhaustive process. Similarly to pair discovery, at each timestamp $t$ (Line 5) we obtain only the values of time series contained in adjacent bins (Lines 6--7). Then, each such value is considered the origin of a search range $\epsilon$, which returns a candidate group at time $t$ (Line 9). Of course, such a candidate group may have been already included in the result bundles previously, during a horizontal expansion at a previous timestamp. In this case, its examination is skipped. Otherwise, if this group contains more than $\mu$ members, we proceed to verify it as a candidate bundle over subsequent timestamps via horizontal expansion (Lines 10-12). As will see next, this expansion may return one or more candidate bundles; each one is checked against the duration constraint $\delta$ before adding it to the result bundles (Lines 13-16).

Regarding the verification step of a candidate bundle $\mathcal{G}^t$, we apply its horizontal expansion over all subsequent timestamps (Line 19). For each member of such candidate bundle (Lines 21-28), we find the group $\mathcal{G}^{t'}$ of time series having values within range $\epsilon$ at time $t'$. Many such new groups may be created, as each member may yield one group. Such a group $\mathcal{G}^{t'}$ may become a new candidate bundle if it satisfies the $\mu$ constraint; if so, it is added to the resulting bundles with an updated duration (Lines 24--27). As we look at subsequent timestamps, it may happen that the same bundle may be added to the results multiple times, but with increasing duration. In the end, we eliminate duplicates and only keep the one with the longest duration (Line 28). Expansion stops once no new candidate bundles can be found in the next timestamp (Lines 29--30).



\setlength{\textfloatsep}{4pt}
\begin{algorithm}[ht!]
	\DontPrintSemicolon
    \KwIn{Set $\mathcal{X}$ of co-evolving time series of length $k$}
	\Parameter{Threshold $\epsilon$, min duration $\delta$, min members $\mu$}
	\KwOut{A list $P$ containing all the discovered bundles}
	\vspace{4pt}
	
	\begin{footnotesize}
	$B \leftarrow CreateBins(\mathcal{X})$ \\
	$P \leftarrow DiscoverBundles(\emptyset, B, \{0, ..., k\}, \epsilon, \delta, \mu)$ \\
	\KwRet $P$
	
	\vspace{4pt}
	\SetKwProg{procFilter}{Procedure}{}{}
	\procFilter{$DiscoverBundles(P, B, T, \epsilon, \delta, \mu)$}{ 
	    \ForEach{$t \in T$}{
            \ForEach{$z \to B.size$}{
                $\mathcal{X}^t \leftarrow B^t_z \cup B^t_{z+1}$ \\
        	    \ForEach{$X \in \mathcal{X}^t$}{
        	        $\mathcal{G}^t \leftarrow getAdjacent(X^t, \epsilon)$ \\
        	        \If{$\mathcal{G}^t \notin P$} {
        	            \If{$\mathcal{G}^t.size \geq \mu$} {
        	                $\mathcal{B} \leftarrow VerifyBundle(\mathcal{G}^t, \{t, .., k\}, \epsilon, \mu)$ \\
        	                \ForEach{$\mathcal{G} \in \mathcal{B}$}{
        	                    \If{$\mathcal{G}.end-\mathcal{G}.start \geq \delta$}{
        	                        $P \leftarrow P \cup \mathcal{G}$
        	                    }
        	                }
        	            }
        	        }
        	    }
    	    }
	    }
	    \KwRet $P$ \\
	}
	
	\vspace{4pt}
	\SetKwProg{procVerify}{Procedure}{}{}
	\procVerify{$VerifyBundle(\mathcal{G}^t, \{t_1, ..., t_\phi\}, \epsilon, \mu, f)$}{
	    $P \leftarrow \mathcal{G}^t$ \\
	    \ForEach{$t'=t_2 \to t_\phi$}{
	        $expanded \leftarrow False$ \\
	        \ForEach{$\mathcal{G} \in P$}{
                \ForEach{$X \in \mathcal{G}$}{
                    $\mathcal{G}^{t'} \leftarrow getAdjacent(X^{t'}, \epsilon)$ \\
                    \If{$(\mathcal{G}^{t'}.size \geq \mu$} {
                        Rearrange duration of $\mathcal{G}^{t'}$ accordingly\\
                        $P \leftarrow P \cup \mathcal{G}^{t'}$ \\
                        $expanded \leftarrow True$ \\
                    }
                }
                $P \leftarrow keepLongestBundles(P)$
	        }
            \If{$expanded=False$}{
                $break$
            }
	    }
		\KwRet $P$ \\
	}
	\end{footnotesize}
	\caption{Sweep line scan bundle discovery}
	\label{alg:simple_scan_bundle}
\end{algorithm}

\begin{algorithm}[ht!]
	\DontPrintSemicolon
	\KwIn{Set $\mathcal{X}$ of co-evolving time series of length $k$}
	\Parameter{Threshold $\epsilon$, min duration $\delta$, min members $\mu$}
	\KwOut{A list $P$ containing all the discovered bundles}
	\vspace{4pt}
	
	\begin{footnotesize}
	$B \leftarrow CreateBins(\mathcal{X})$ \\
	$C \leftarrow GetCheckpoints(k, \delta)$ \\
	$P \leftarrow DiscoverBundles(\emptyset, B, C, \epsilon, \delta, \mu)$ \\
	\KwRet $P$ \\
	
	\vspace{4pt}
	\SetKwProg{procVerify}{Procedure}{}{}
	\procVerify{$VerifyBundle2Way(\mathcal{G}^t, t, k, \epsilon, \mu)$}{
	    $P, Q_1, Q_2 \leftarrow \emptyset$ \\
	    $Q_1 \leftarrow VerifyBundle(\mathcal{G}^t, \{t-1, ..., 0\}, \epsilon, \mu)$ \\
	    \ForEach{$q \in Q_1$}{
	        $t_c \leftarrow q.start + \delta - 1$ \\
	        $Q_2 \leftarrow Q_2 \cup VerifyBundle(q, \{t_c, ..., q.{start}\}, \epsilon, \mu)$ \\
	    }
	    \ForEach{$q \in Q_2$}{
            \If{$q.end - q.start \geq \delta$}{
                $P \leftarrow P \cup VerifyBundle(q, \{t_c, ..., k\}, \epsilon, \mu)$ \\
            }	    
	    }
		\KwRet $P$ \\
	}
	\end{footnotesize}
	\caption{Checkpoint scan bundle discovery}
	\label{alg:checkpoint_scan_bundle}
\end{algorithm}

\subsection{Bundle Discovery Using Checkpoints}
\label{sec:checkpoint_bundle}
For bundle discovery using checkpoints, we apply a similar sweep line approach, but this time we only filter at checkpoints and then verify towards both directions in the time axis. Algorithm \ref{alg:checkpoint_scan_bundle} describes this procedure. After initializing the checkpoints (Line 2), we run the bundle discovery process as in Algorithm \ref{alg:simple_scan_bundle}, but this time looking at checkpoints (Line 3) instead of all timestamps. For the two-way horizontal expansion, we first examine all candidate bundles in set $Q_1$ detected from the current checkpoint towards the origin of time axis (Line 7). This is done because a qualifying bundle could have started earlier, before the current checkpoint. Afterwards, we apply the same eager pruning strategy as in Section~\ref{subsec:checkpoint_join}. So, we verify each such candidate bundle jumping forward at timestamp $t_c = q.start + \delta$ and continue backwards in time to its currently known start (Lines 8--10). Among the candidate bundles (in set $Q_2$) returned from the forward verification (Line 11), we care only for those that satisfy the minimal duration constraint $\delta$ (Line 12). These are further verified from timestamp $t_c$ and forward in time, obtaining all subsequent qualifying bundles (Line 13).

\section{Experimental Evaluation}
\label{sec:eval}

\subsection{Experimental Setup}
\label{subsec:evaluation_setup}
We evaluate the performance of our methods on pair and bundle discovery both qualitatively and quantitatively. We compare our checkpoint (CP) scan approaches for each problem with the respective sweep line (SL) methods. We use a real-world and a synthetic dataset as listed in Table~\ref{tab:datasets}. Next, we describe their characteristics.

\begin{table}[t]
\caption{Datasets used in the experiments.}
\centering
\begin{small}
\begin{tabular}{lcc} 
\hline
{\em Dataset} & {\em Size} & {\em Time series length} \\
\hline
Water & 822 & 168  \\
Synthetic & 50,000 & 1,000 \\
\hline
\end{tabular}
\end{small}
\label{tab:datasets}
\end{table}

\emph{DAIAD Water Consumption (Water)}. Courtesy of the DAIAD project\footnote{ \url{http://daiad.eu/}}, we acquired a time series dataset of hourly water consumption for 822 households in Alicante, Spain from 1/1/2015 to 20/1/2017. In order to get a more representative dataset for our tests, we first calculated the weekly time series ($24 \times 7$ timestamps, one value per hour from Monday to Sunday) per household by averaging corresponding hourly values over the entire period.

\emph{Synthetic Dataset}. We generated a synthetic dataset of 50,000 time series, each with a length of 1,000 timestamps. So, this dataset contains 50 million data points in total. The dataset was generated in a similar manner to the synthetic dataset used in \cite{keogh1999indexing}.

All experiments were conducted on a Dell PowerEdge M910 with 4 Intel Xeon E7-4830 CPUs, each containing 8 cores clocked at 2.13GHz, 256 GB RAM and a total storage space of 900 GB.

\subsection{Evaluation Results}
\label{subsec:exp_results}
We conducted two sets of experiments, using the water and synthetic datasets respectively. The water dataset was used for qualitative and quantitative assessment of our methods on pair and bundle discovery, while the synthetic dataset was used to evaluate their efficiency in terms of execution time.

\subsubsection{Pair and Bundle Discovery over Real Data}
\label{subsubsec:pair_bundle}
We performed several experiments using the water dataset with various parameter values to detect pairs and bundles using both the SL and CP approaches. The water dataset was \textit{$z$-normalized} in order to eliminate larger amplitude discrepancies among the time series and focus on structural similarity.



To evaluate our methods against different parameters, we performed preliminary tests to extract ranges of values where the algorithms would return a reasonable number of results. Table \ref{tab:parameters1} lists the range of values for the parameters used for bundle and pair discovery tests (recall that parameter $\mu$ is not applicable in pair discovery); default values are in bold. Parameter $\delta$ is expressed as a percentage of the duration of the time series, $\epsilon$ is expressed as a percentage of the value range (i.e., difference $max-min$ in values encountered across the dataset) and $\mu$ is expressed as a percentage of the number of time series in the dataset.

\begin{table}[t]
\caption{Parameters for tests over the water dataset}
\centering
\begin{small}
\begin{tabular}{lc} 
\hline
{\em Parameter} &{\em Values} \\
\hline
$\delta$ (\% of time series length, i.e., 168) & 6\%, 5\%, \textbf{6\%}, 7\%, 8\% \\
$\epsilon$ (\% of value range, i.e., approx 11.4) & 4\%, 5\%, \textbf{6\%}, 7\%, 8\% \\
$\mu$ (\% of dataset size, i.e., 822) & 0.5\%, 0.75\%, \textbf{1\%}, 1.25\%, 1.5\% \\
\hline
\end{tabular}
\end{small}
\label{tab:parameters1}
\end{table}

\paragraph{Varying $\delta$}
\begin{figure*}[!ht]
 \centering
 \subfigure[Bundle discovery exec time]{\includegraphics[trim=0.5cm 0.5cm 0.5cm 0.5cm, clip, width=0.246\textwidth]{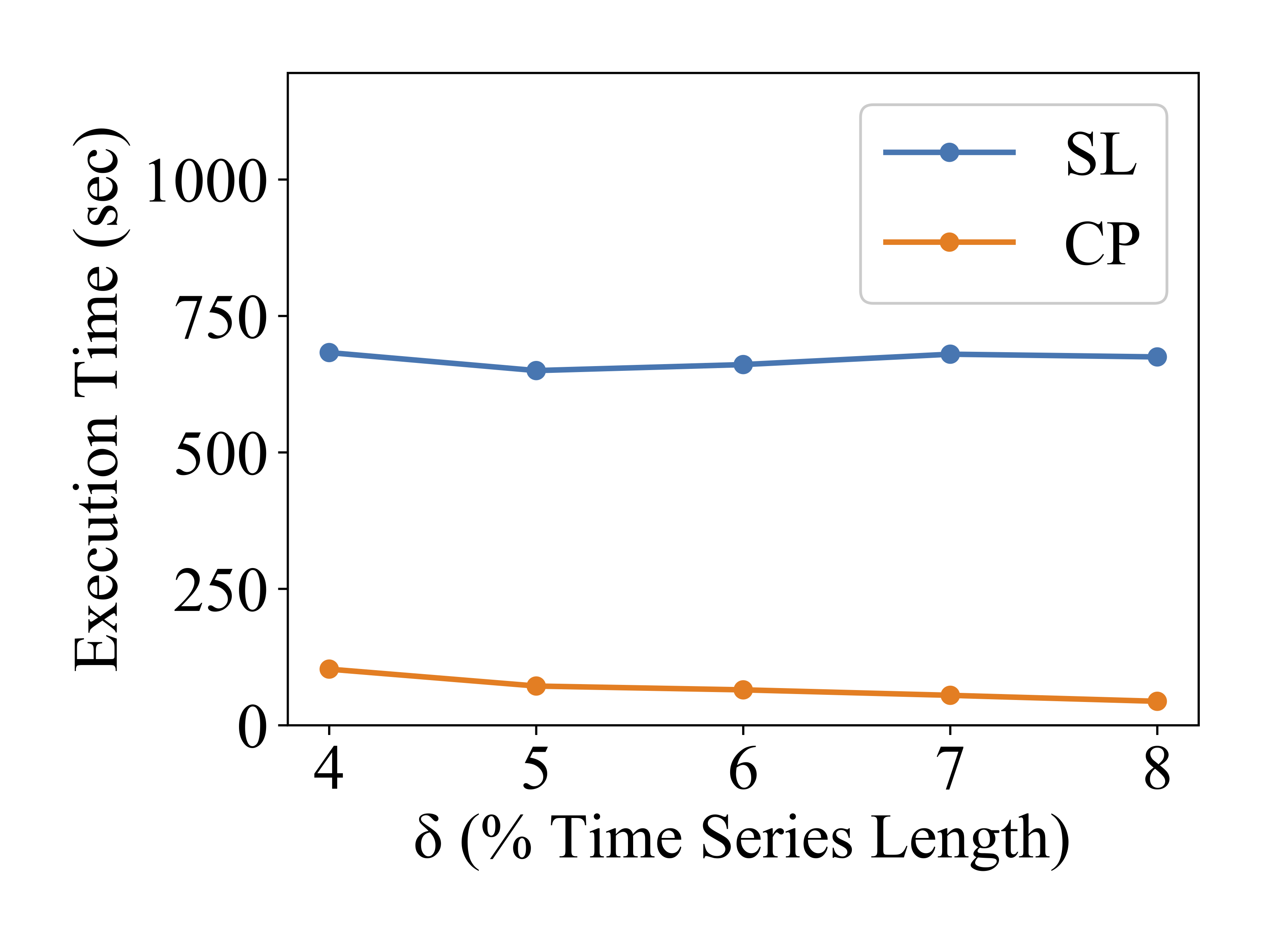}\label{subfig:var_delta}}
 \subfigure[Bundle discovery results]{\includegraphics[trim=0.5cm 0.5cm 0.5cm 0.5cm, clip,width=0.246\textwidth]{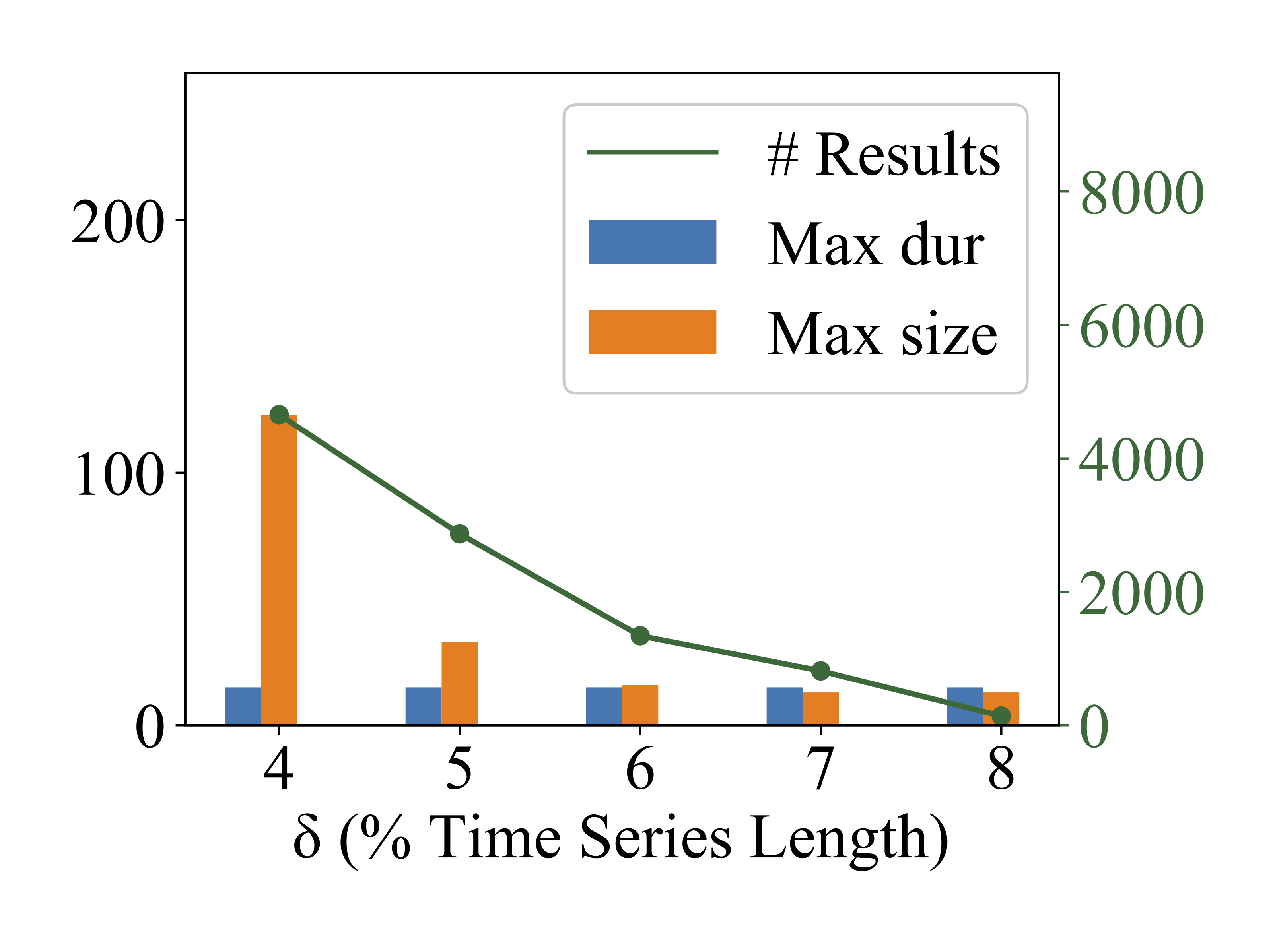}\label{subfig:var_delta_bars}}
 \subfigure[Pair discovery exec time]{\includegraphics[trim=0.5cm 0.5cm 0.5cm 0.5cm, clip,width=0.246\textwidth]{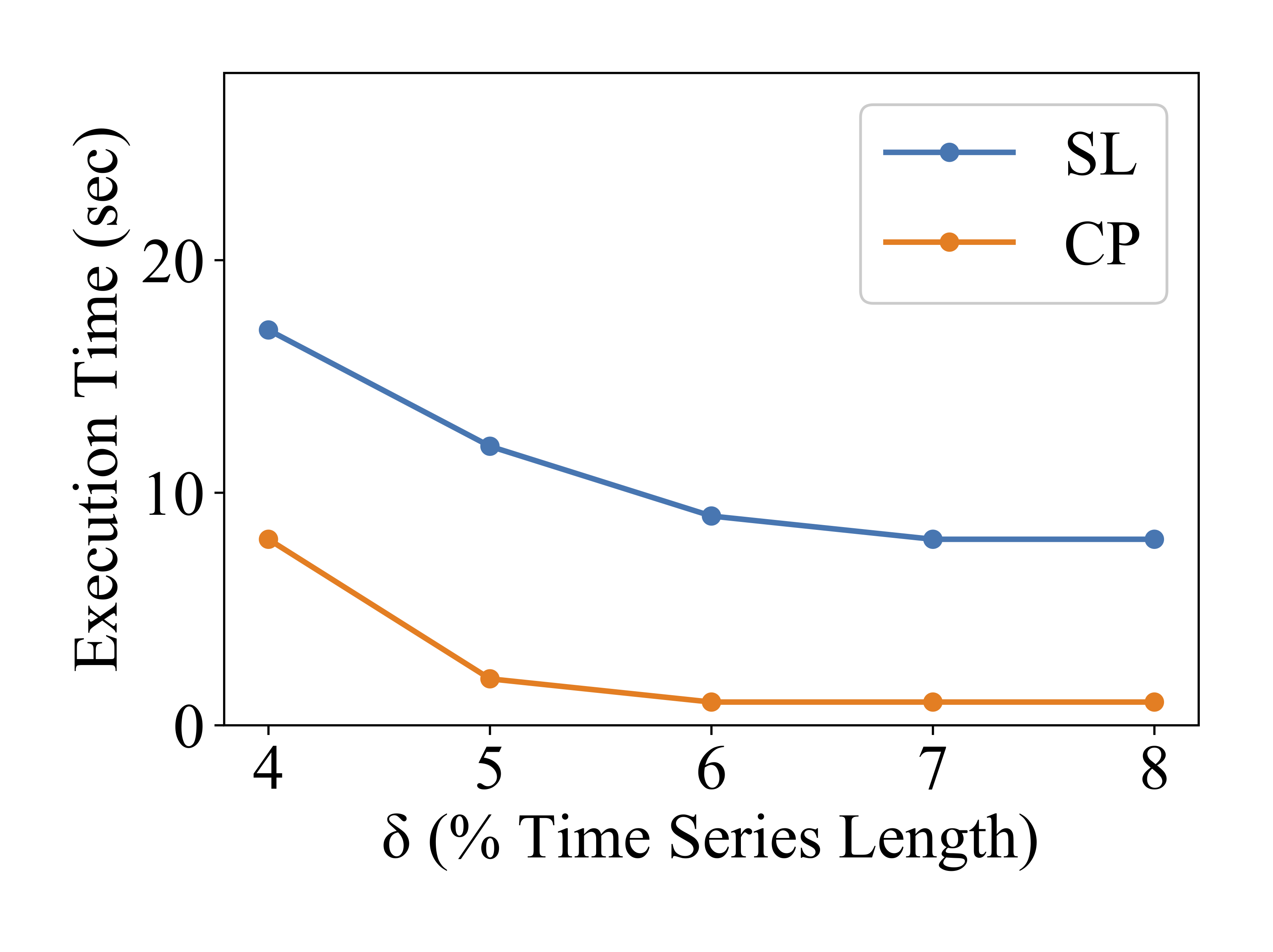}\label{subfig:var_delta_bars_pairs}}
 \subfigure[Pair Discovery results]{\includegraphics[trim=0.5cm 0.5cm 0.5cm 0.5cm, clip,width=0.246\textwidth]{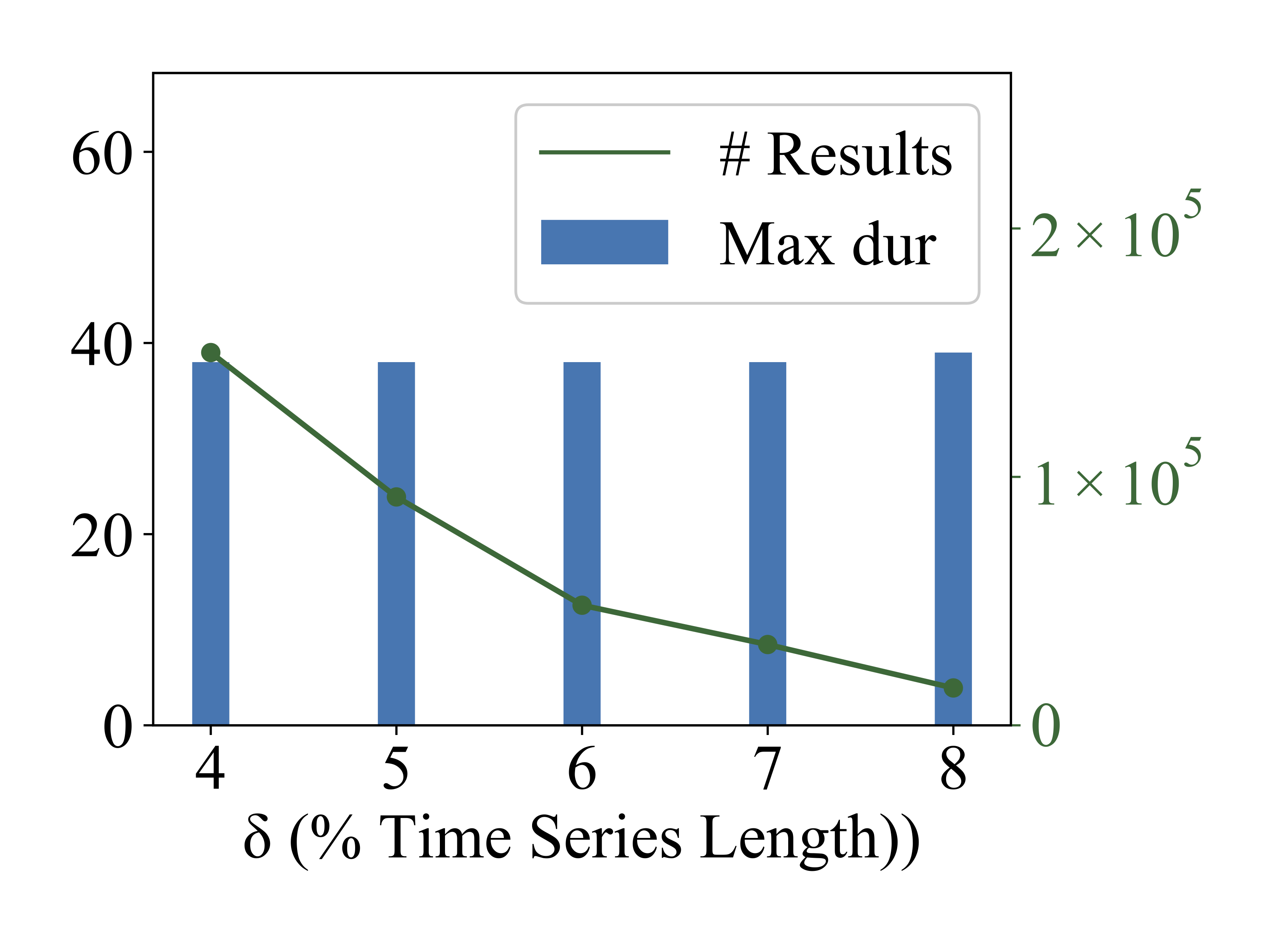}\label{subfig:var_delta_bars_pairs_bars}}
 \caption{Assessment against real data for varying $\delta$.}
 \label{fig:exp1}
\end{figure*}

Figure \ref{fig:exp1} depicts the results by varying the minimal duration $\delta$. In bundle discovery, the CP algorithm outperforms SL up to an order of magnitude in terms of execution time (Figure \ref{subfig:var_delta}). As the threshold $\delta$ gets larger, performance is further improved as less checkpoints are specified and thus less candidates need verification. On the contrary, the SL approach performs similarly irrespective of $\delta$, as time series must be checked at all timestamps. From Figure \ref{subfig:var_delta_bars}, it turns out that the number of detected bundles is reduced as the $\delta$ value increases, which is expected as fewer bundles can last longer. In this plot, the blue bars indicate the maximum bundle duration among the ones that were detected, while the orange bars indicate the larger detected bundle in terms of membership. It is clear that the maximum bundle size is drastically reduced as the number of results diminish, while the maximum duration among bundles remains the same with the increase of $\delta$, because the longest bundle is the same in these results.

Regarding pair discovery, since it is an overall faster process, the differences in terms of efficiency are smaller, but still apparent. In this case, the execution time (Figure \ref{subfig:var_delta_bars_pairs}) is more abruptly reduced in both SL and CP methods, since less subsequences qualify as pairs. The number of results (Figure \ref{subfig:var_delta_bars_pairs_bars}) is now naturally much larger, as far more pairs are expected to be verified if bundles exist. The same stands for the maximum duration among pairs, which tend to last longer compared to bundles. Since a pair is actually a bundle with $\mu$=2 members, it is easier to find local similarity over longer intervals between two subsequences rather than an increased number of them. The maximum duration, as in bundle discovery, remains the same as $\delta$ increases, since this corresponds to the same pair in the results. 

\paragraph{Varying $\epsilon$}
\begin{figure*}[!ht]
 \centering
 \subfigure[Bundle discovery exec time]{\includegraphics[trim=0.5cm 0.5cm 0.5cm 0.5cm, clip,width=0.246\textwidth]{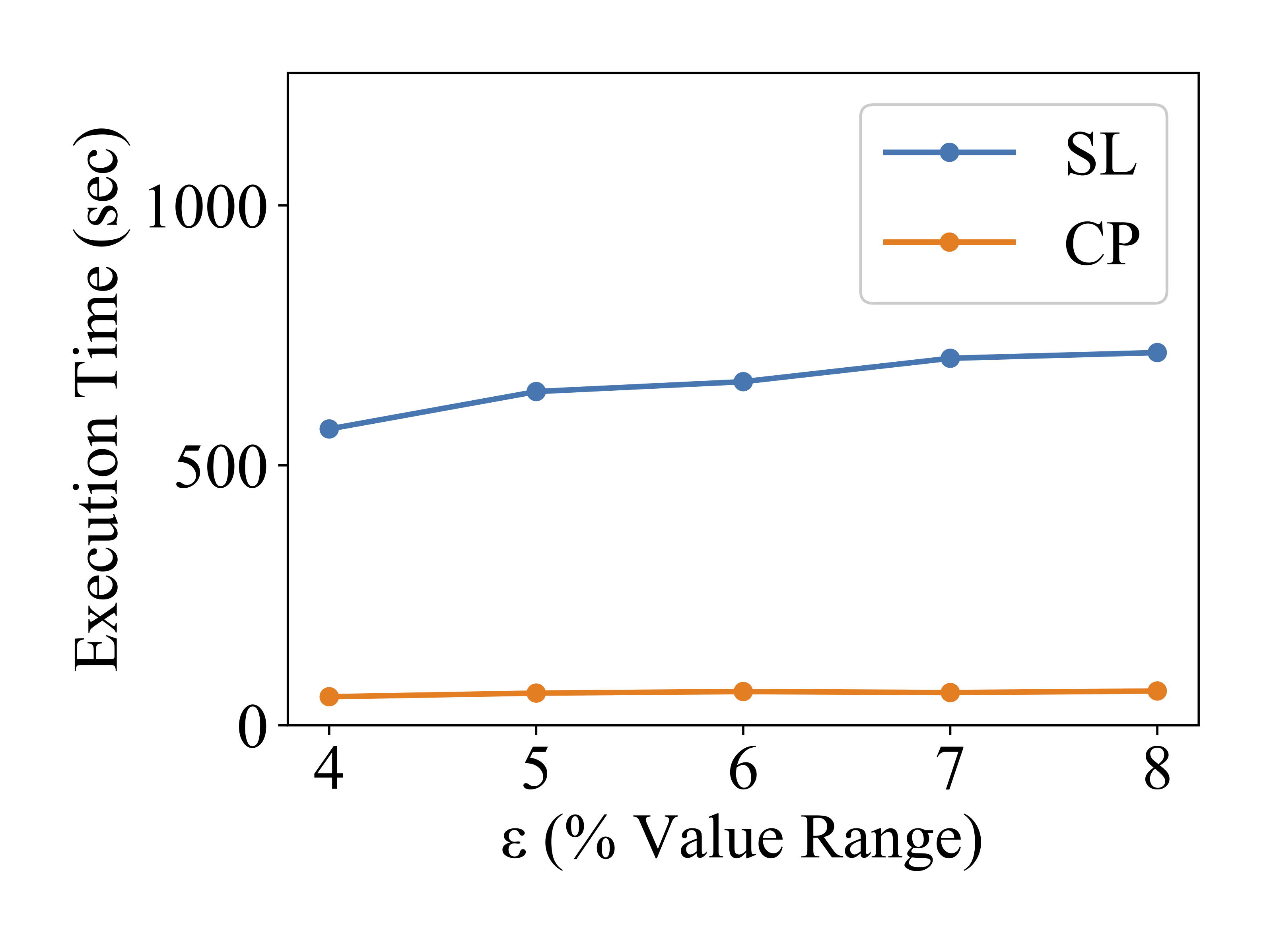}\label{subfig:var_eps}}
 \subfigure[Bundle discovery results]{\includegraphics[trim=0.5cm 0.5cm 0.5cm 0.5cm, clip,width=0.246\textwidth]{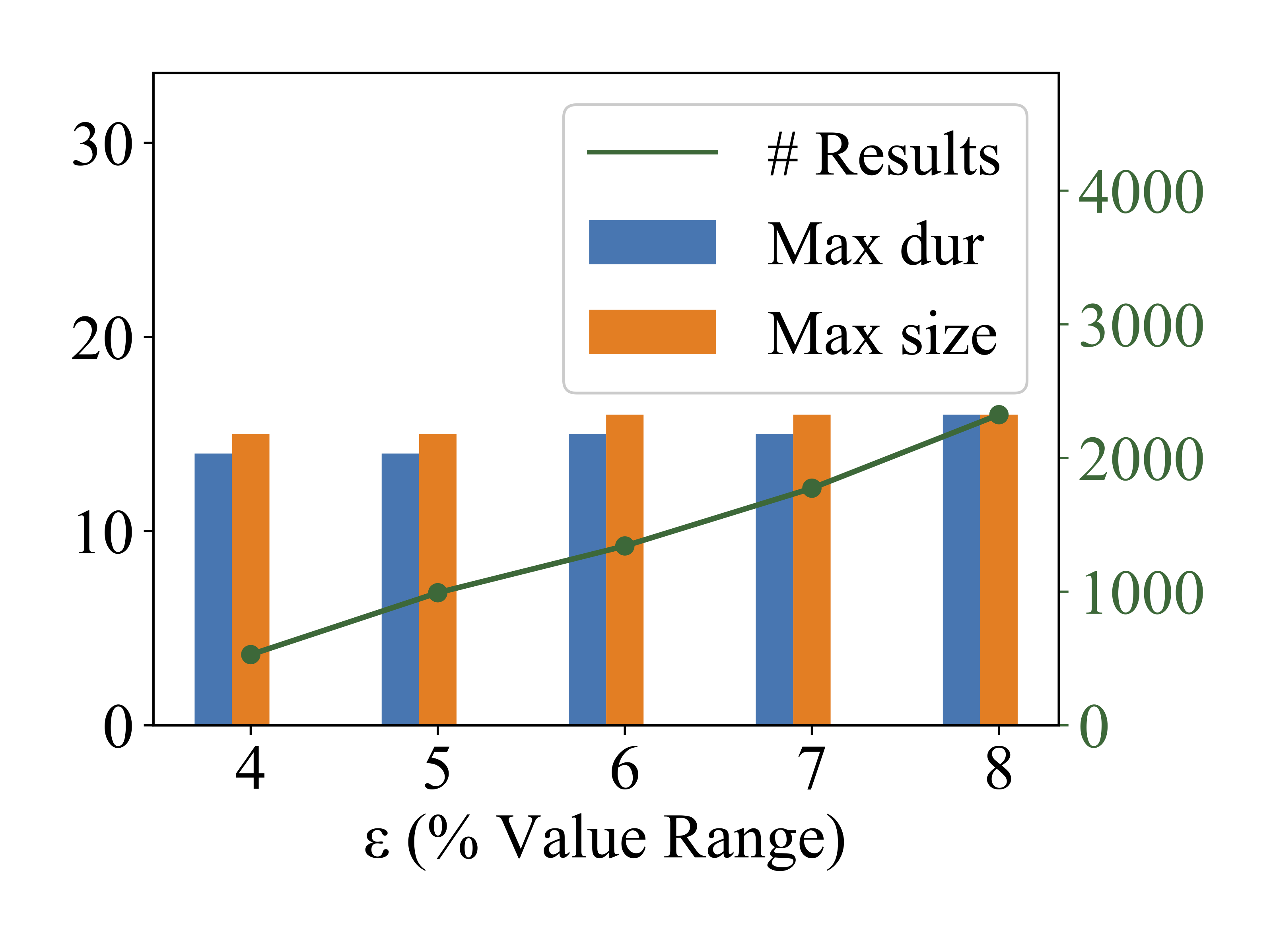}\label{subfig:var_eps_bars}}
 \subfigure[Pair discovery exec time]{\includegraphics[trim=0.5cm 0.5cm 0.5cm 0.5cm, clip,width=0.246\textwidth]{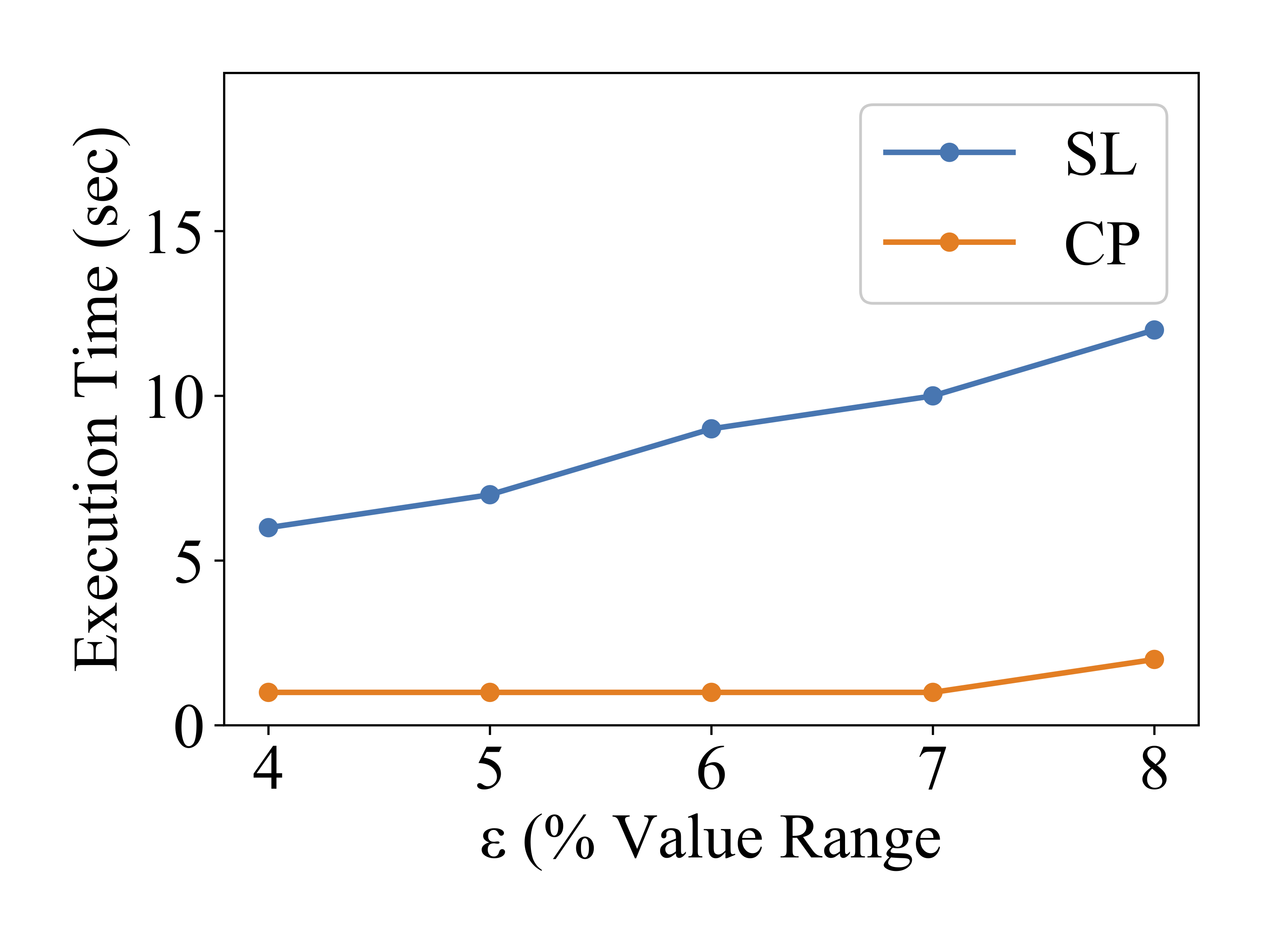}\label{subfig:var_eps_bars_pairs}}
 \subfigure[Pair Discovery results]{\includegraphics[trim=0.5cm 0.5cm 0.5cm 0.5cm, clip,width=0.246\textwidth]{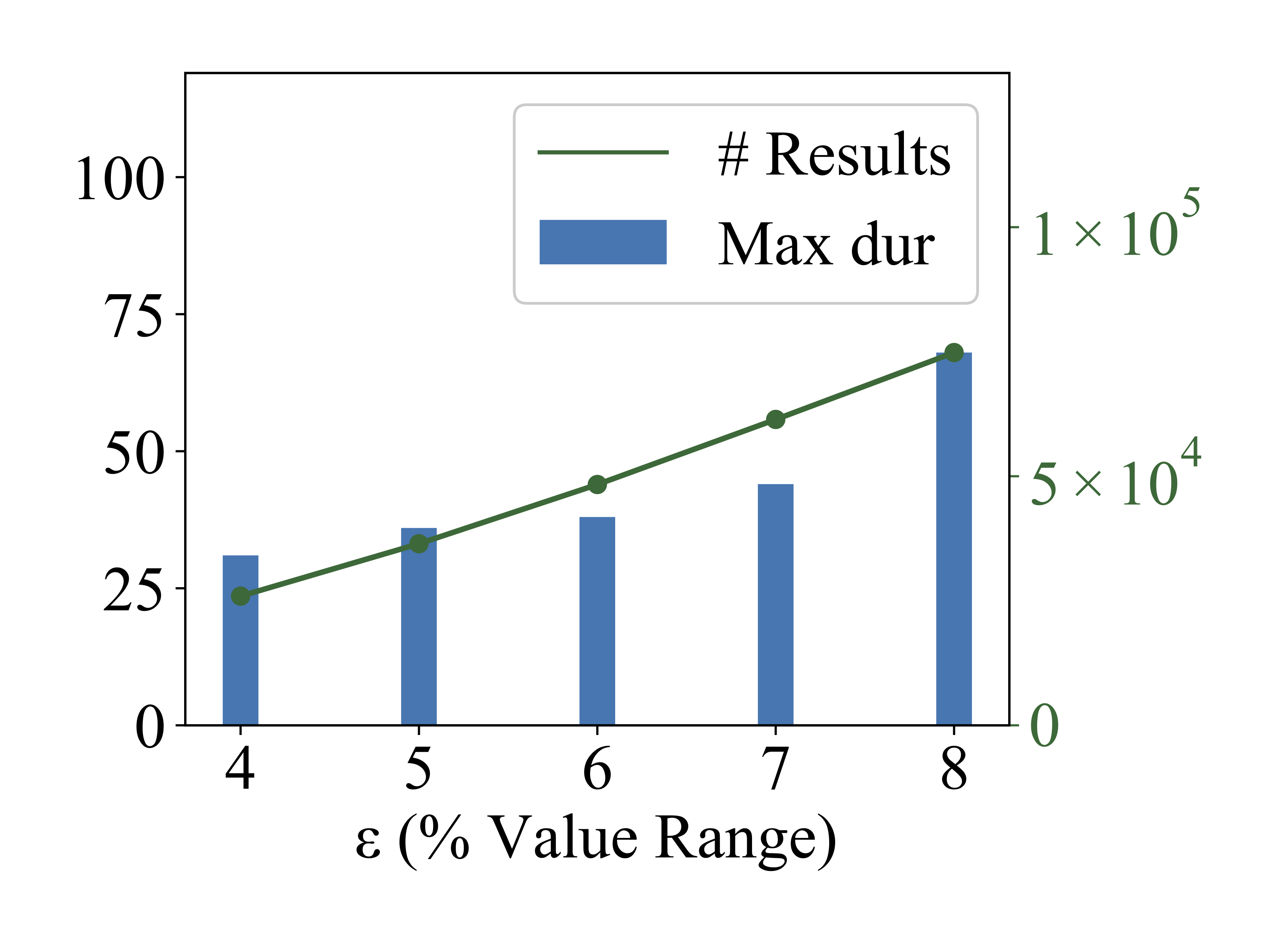}\label{subfig:var_eps_bars_pairs_bars}}
 \caption{Assessment against real data for varying $\epsilon$.}
 \label{fig:exp2}
\end{figure*}

Varying threshold $\epsilon$ for bundle discovery slightly incurs more execution cost for both SL and CP approaches. This is due to the increased number of bundles that need to be verified. Nonetheless, the difference in cost remains at levels of at most an order of magnitude, as shown in Figure \ref{subfig:var_eps}. As expected, the number of results is also increased (Figure \ref{subfig:var_eps_bars}. So does the maximum duration bundle, which is also expected due to more qualifying bundles, hence a higher probability to find longer ones. The maximum bundle size (i.e., membership) is also increased, as more time series can form a bundle when allowing a wider threshold $\epsilon$ in deviation of their respective values.

Regarding pair discovery, the results are again similar to bundle discovery for varying $\epsilon$. For very small $\epsilon$ values of up to 7\% of the value range, the CP algorithm returns results almost instantly. The SL approach is at least five times slower, with its performance deteriorating more rapidly with increasing $\epsilon$ values. Again, as in bundle discovery, the results are growing with greater $\epsilon$, as does the maximum duration among pairs, especially for $\epsilon$ equal to 8\% of the value range.

\paragraph{Varying $\mu$}
\begin{figure}[!ht]
 \centering
 \subfigure[Bundle discovery execution time]{\includegraphics[trim=0.5cm 0.5cm 0.5cm 0.5cm, clip,width=0.4\textwidth]{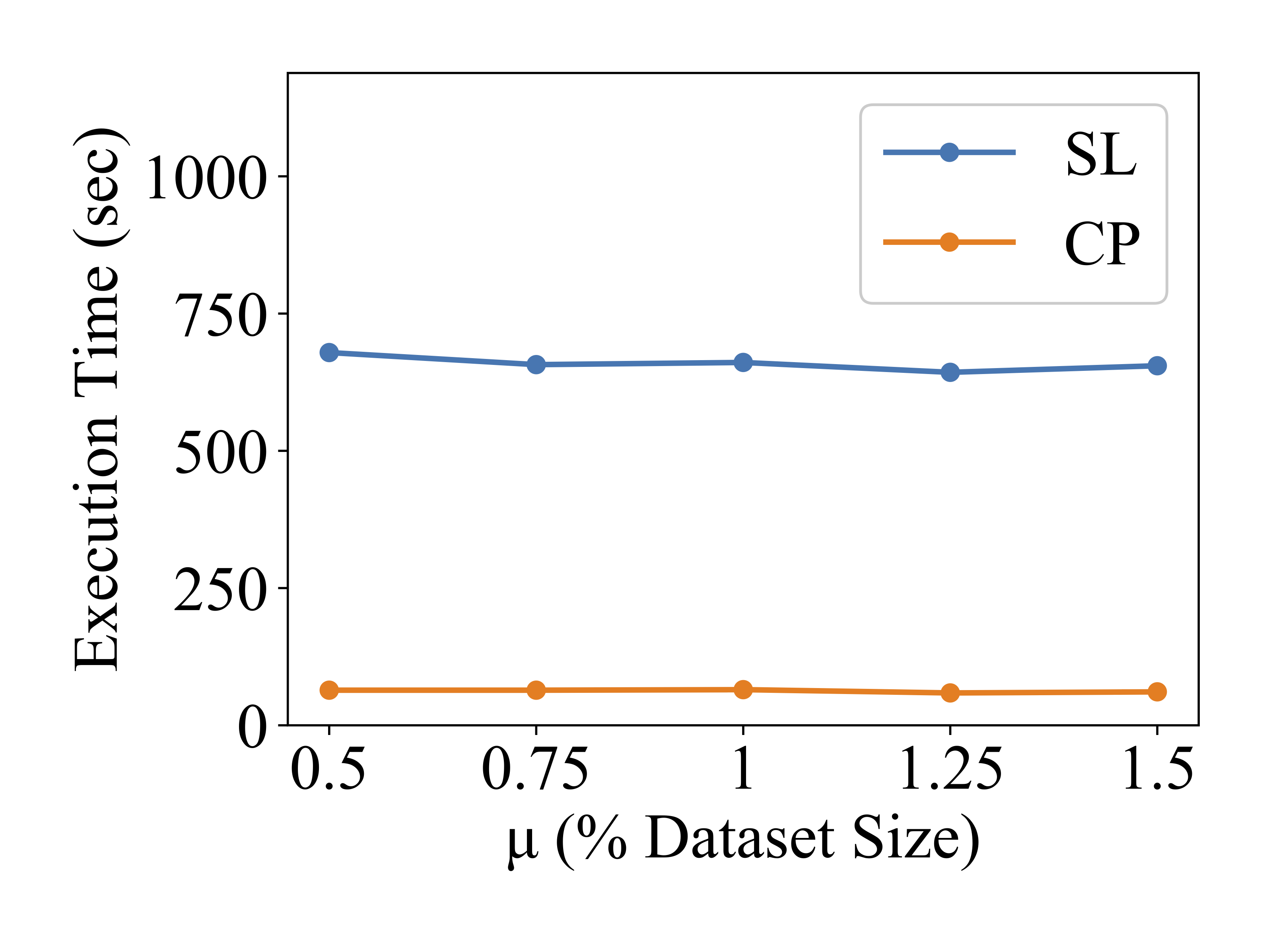}\label{subfig:var_mu}}
 \subfigure[Bundle discovery results]{\includegraphics[trim=0.5cm 0.5cm 0.5cm 0.5cm, clip,width=0.4\textwidth]{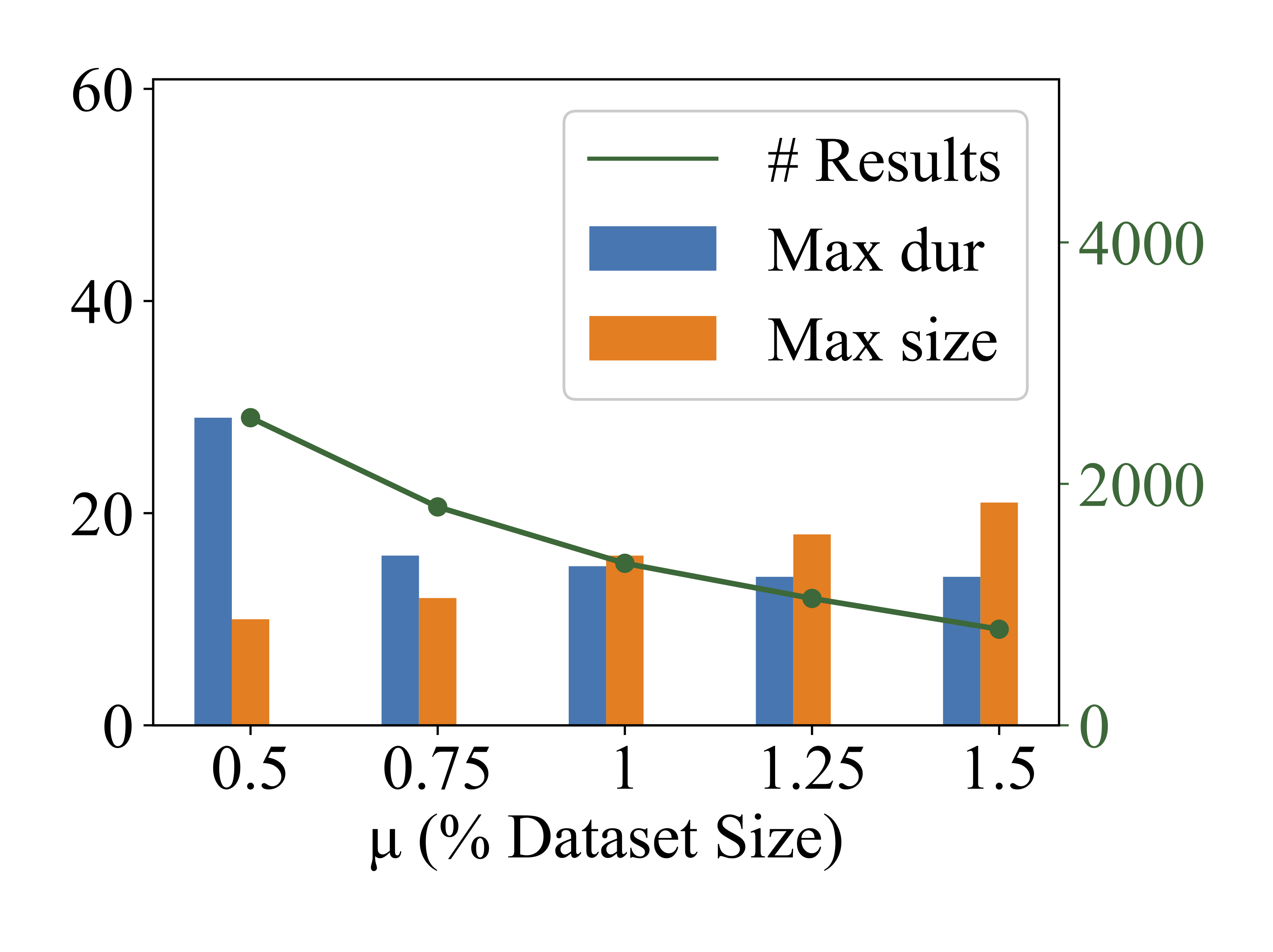}\label{subfig:var_mu_bars}}
 \caption{Assessment against real data for varying $\mu$.}
 \label{fig:exp3}
\end{figure}

When varying the minimum membership parameter $\mu$ in bundle discovery (Figure \ref{fig:exp3}), the results regarding execution time are again very similar to the rest of the tests as indicated in Figure \ref{subfig:var_mu}. Again, the CP algorithm outperforms SL up to one order of magnitude. The execution time is very slightly decreased for larger $\mu$ values in both algorithms, as more candidate bundles are pruned. Results from CP are reported almost instantly, since the number of time series is rather small and scanning through the limited number of checkpoints is very fast. This explains why performance of SL does not get drastically improved as $\mu$ gets larger, since filtering and verification has to be repeated at every timestamp. The number of results (Figure~\ref{subfig:var_mu_bars}) is reduced as $\mu$ increases, which is expected, as less bundles get detected with a larger membership. The maximum duration among bundles also decreases; interestingly, the maximum size detected among bundles increases as the number of results diminishes, due to the growing number $\mu$ of required number of members per bundle.

\subsubsection{Efficiency against Synthetic Data}
\label{subsubsec:effic}
To evaluate the efficiency of our methods, we used the synthetic dataset. Regarding parameter values, as in the previous experiments, we performed preliminary tests to extract ranges of values where the algorithms return a reasonable number of results. Table \ref{tab:parameters2} lists the range of values for all parameters used in these efficiency tests for bundle and pair discovery, with the default values emphasized in bold (again, $\mu$ is not applicable in pair discovery).


\paragraph{Varying Dataset Size}
\begin{figure}[tb]
 \centering
 \subfigure[Bundle discovery]{\includegraphics[trim=0.5cm 0.5cm 0.5cm 0.5cm, clip,width=0.4\textwidth]{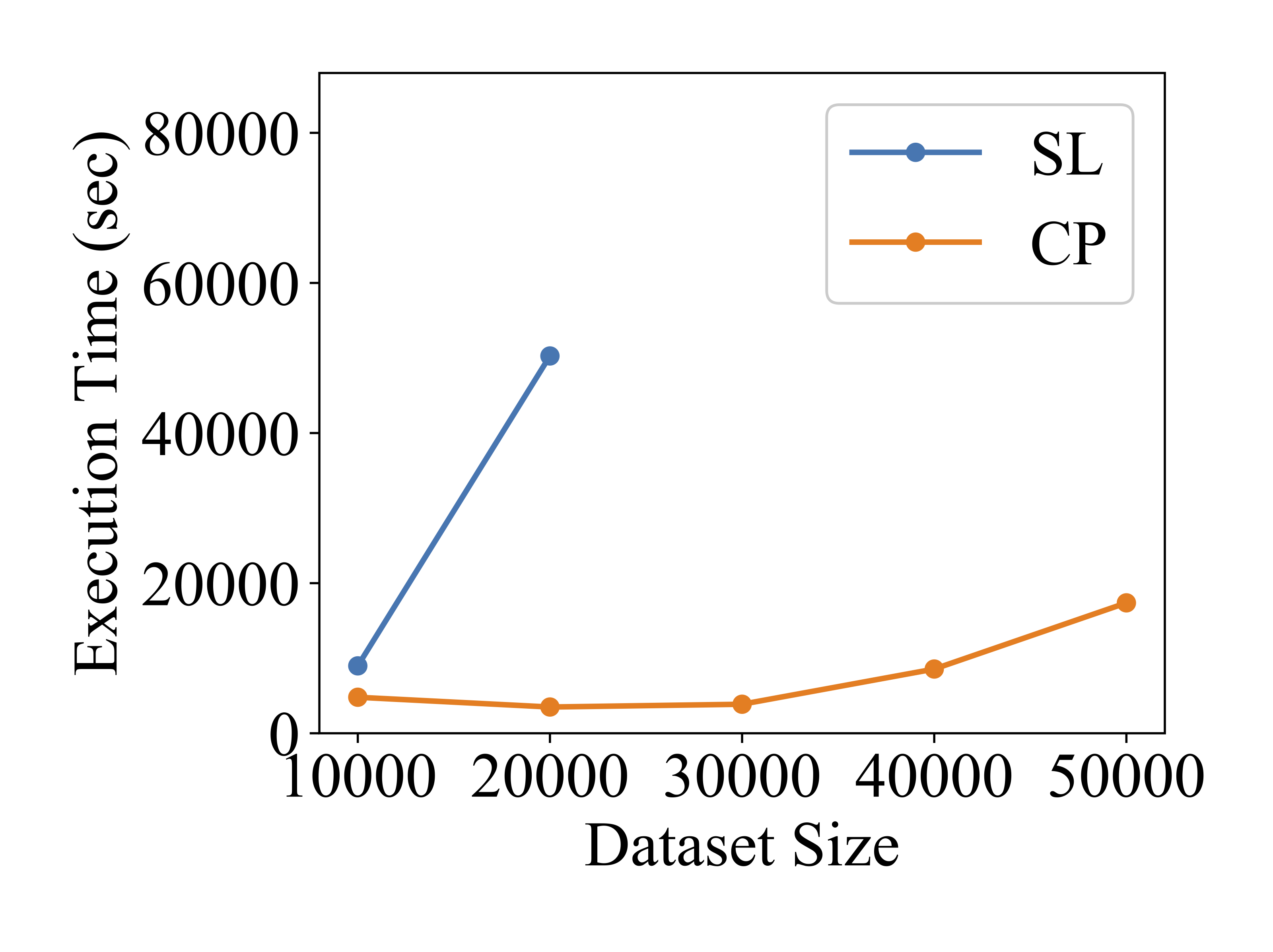}\label{subfig:var_dat_size}}
 \subfigure[Pair discovery]{\includegraphics[trim=0.5cm 0.5cm 0.5cm 0.5cm, clip,width=0.4\textwidth]{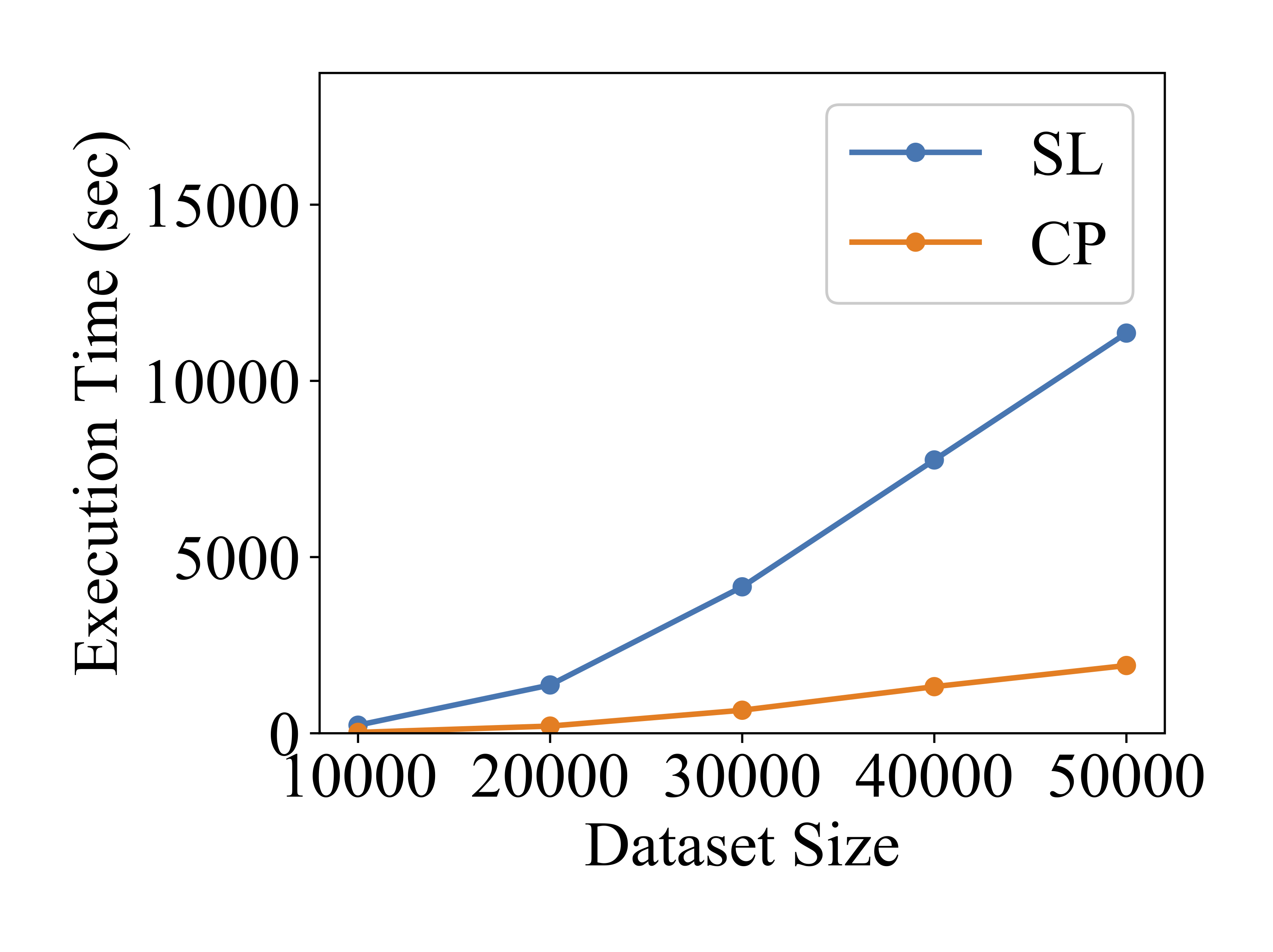}\label{subfig:var_dat_size_pairs}}
 \caption{Efficiency with varying numbers of time series.}
 \label{fig:exp4}
\end{figure}

\begin{table}[tb]
\caption{Parameters for tests against synthetic data}
\centering
\begin{small}
\begin{tabular}{lc} 
\hline
{\em Parameter} &{\em Values} \\
\hline
Dataset Size & \textbf{10000}, 20000, 30000, 40000, 50000 \\
Time Series Length & 600, 700, \textbf{800}, 900, 1000 \\
$\delta$ (\% of time series length) & \textbf{2.5\%} \\
$\epsilon$ (\% of value range) & \textbf{0.2\%} \\
$\mu$ (\% of dataset size) & \textbf{1.25\%} \\
\hline
\end{tabular}
\end{small}
\label{tab:parameters2}
\end{table}

Figure \ref{fig:exp4} depicts the performance comparison between CP and SL algorithms for bundle and pair discovery. We omit cases where execution of an algorithm was taking more than 15 hours (cutoff). As illustrated in Figure \ref{subfig:var_dat_size}, an increase in the dataset size leads to a very abrupt deterioration of performance for the SL algorithm of up to several hours of execution for 20,000 time series. For larger dataset sizes, the execution time was significantly longer than the cutoff time. On the other hand, CP reports results in all cases. Its execution time increases for larger dataset sizes, but manages to finish in a few hours in the worst case (for 50,000 time series). It is worth noting that membership parameter $\mu$ is more relaxed for larger dataset sizes, as it is expressed in terms of percentage of the total number of time series in the dataset. This explains the slight improvement in performance  for dataset sizes of 20,000 and 30,000 for the CP method. In general, CP is more than an order of magnitude faster than SL, which also stands for the case of pair discovery, as illustrated in Figure \ref{subfig:var_dat_size_pairs}. As the number of time series in the dataset grows, it is natural that more pairs will be detected, hence the linear increase in the execution cost for the CP algorithm. Of course, baseline SL requires more time, as it must check many more combinations of time series.

\paragraph{Varying Time Series Length}
\begin{figure}[tb]
 \centering
 \subfigure[Bundle discovery]{\includegraphics[trim=0.5cm 0.5cm 0.5cm 0.5cm, clip,width=0.4\textwidth]{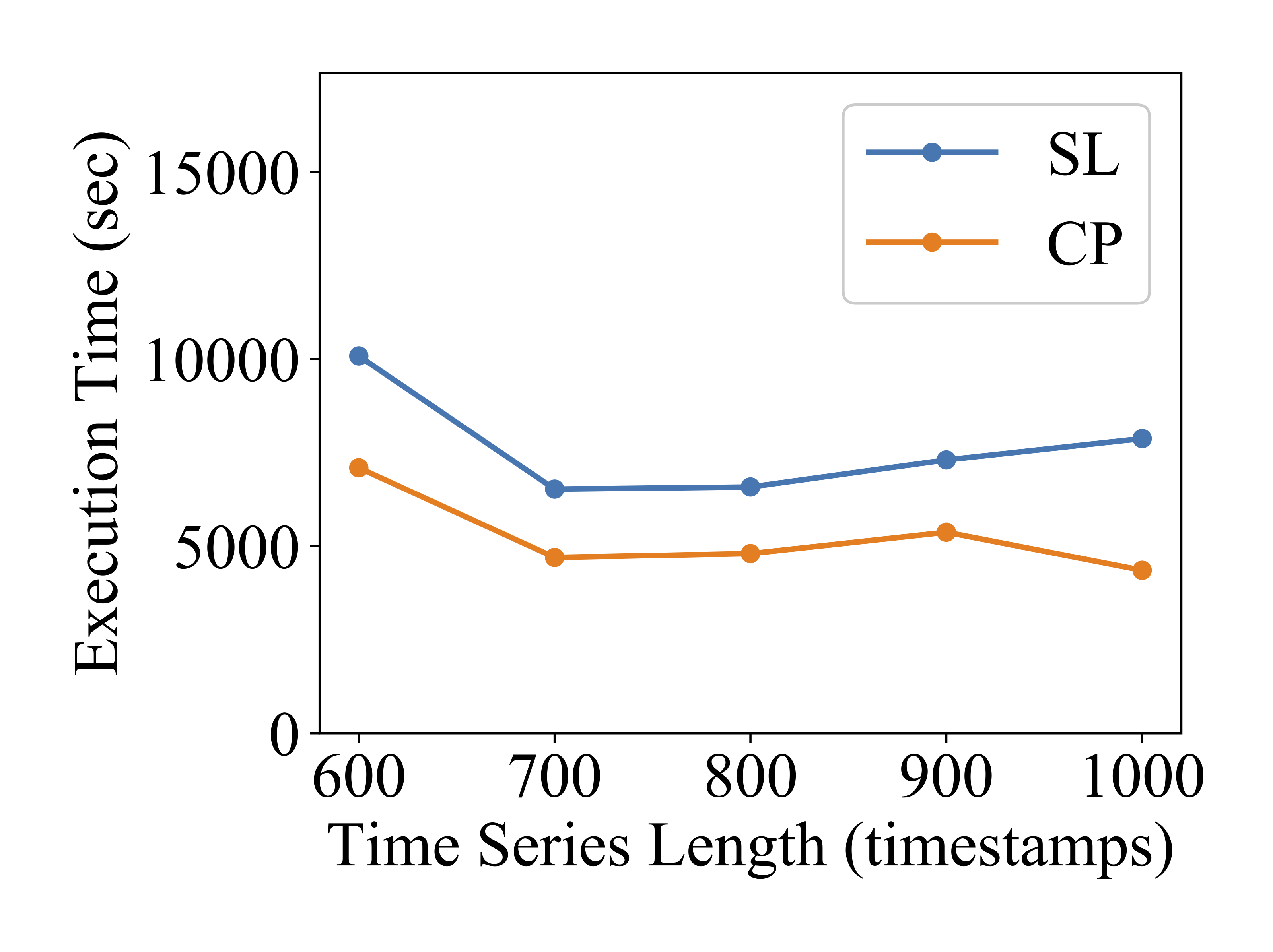}\label{subfig:var_ts_length}}
 \subfigure[Pair discovery]{\includegraphics[trim=0.5cm 0.5cm 0.5cm 0.5cm, clip,width=0.4\textwidth]{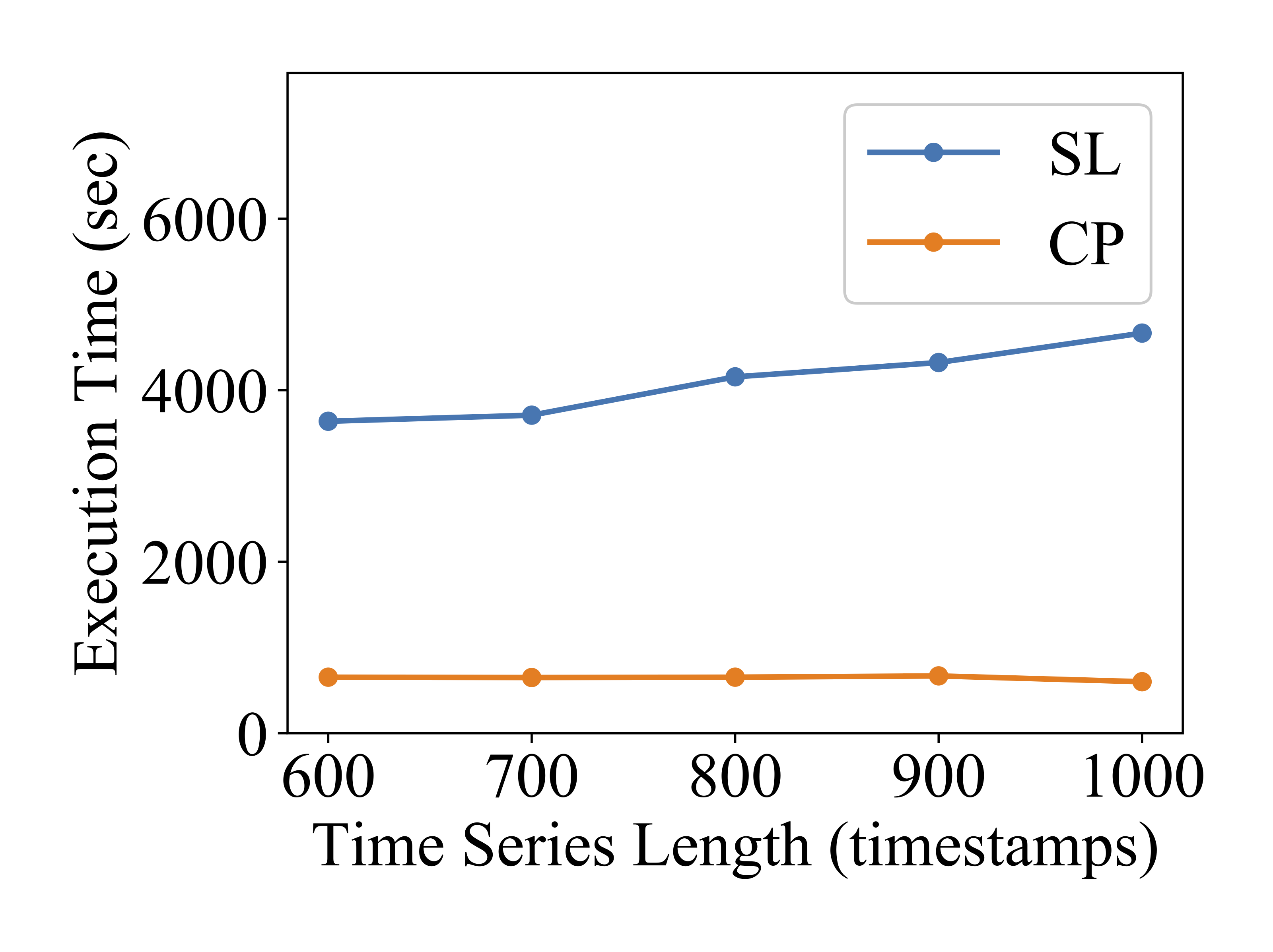}\label{subfig:var_ts_length_pairs}}
 \caption{Efficiency with varying length of time series.}
 \label{fig:exp5}
\end{figure}

For time series with increasing length (Figure~\ref{subfig:var_ts_length}), the CP algorithm for bundle discovery again constantly outperforms SL. Similarly to previous experiments, $\delta$ is expressed as a percentage of the time series length. We observe that the execution time initially decreases for both algorithms, as more bundles are pruned. However, as the time series length (and $\delta$) gets larger, the performance of both algorithms slightly worsens. Only in the case of 1,000 timestamps the execution time starts to drop for the CP algorithm due to the even larger $\delta$. This is not the case with the SL method, which has to evaluate more timestamps. Similar observations stand for pair discovery, with the CP algorithm significantly outperforming SL in all cases (Figure~\ref{subfig:var_ts_length_pairs}).
\section{Conclusions}
\label{sec:conclusions}

In this paper, we addressed the problems of pair and bundle discovery over co-evolving time series, according to their local similarity. We introduced two efficient algorithms for pair and bundle discovery that utilize checkpoints appropriately placed across the time axis in order to aggressively prune the search space and avoid expensive exhaustive search at each timestamp. Our methods successfully detect locally similar subsequences of co-evolving time series, as demonstrated in our experimental evaluation over real-world and synthetic data. Also, they were orders of magnitude faster compared to baseline methods that apply a sweep line approach, confirming their effectiveness in pair and bundle discovery. In the future, we plan to relax the strict conditions regarding $\epsilon$ and $\delta$ thresholds and further improve the scalability of our algorithms to extend their applicability over very large time series datasets, both in terms of cardinality, as well as in terms of length.

\bibliographystyle{abbrv}
\bibliography{references}

\end{document}